\documentclass[sigplan,nonacm]{acmart}

\AtBeginDocument{%
  \providecommand\BibTeX{{%
    \normalfont B\kern-0.5em{\scshape i\kern-0.25em b}\kern-0.8em\TeX}}
    \DeclareSIUnit\bit{b}}

\usepackage{booktabs, multirow} 
\usepackage{soul}
\usepackage{amsmath}
\usepackage{url}
\usepackage{hyperref}
\usepackage{graphicx}
\usepackage{listings}
\usepackage{placeins}
\usepackage{etoolbox}
\usepackage{subcaption}
\usepackage{cleveref}
\usepackage{circledsteps}

\usepackage{xurl}

\usepackage{siunitx}
\usepackage{moreverb}

\hypersetup{colorlinks,
            breaklinks,         
            citecolor=blue,
            linkcolor=blue,
            urlcolor=blue,
}

\newcommand{\isrc}[1]{\lstinline[basicstyle=\ttfamily]|#1|}

\newcommand{\STEP}[1]{\makebox{step\hspace{0.7mm}{\Circled{#1}}}}
\newcommand{\CIRC}[1]{\Circled{#1}}

\begin{document}

\title[Julia Cloud Matrix Machine]
{Julia Cloud Matrix Machine: Dynamic Matrix Language Acceleration on Multicore Clusters in the Cloud}

\newcommand{\commentBlock}[1]{}

\author{Jay~Hwan~Lee}
\email{jlee758@yonsei.ac.kr}
\author{Yeonsoo~Kim}
\email{yeonsoo.kim@yonsei.ac.kr}
\author{Yonghyun~Ryu}
\affiliation{
    \institution{Yonsei University}
    \city{Seoul}
    \country{South Korea}
}
\author{Wasuwee~Sodsong}
\author{Hyunjun~Jeon}
\author{Jinsik~Park}
\author{Bernd~Burgstaller}
\email{bburg@yonsei.ac.kr}
\affiliation{
    \institution{Yonsei University}
    \city{Seoul}
    \country{South Korea}
}
\author{Bernhard~Scholz}
\email{bernhard.scholz@sydney.edu.au}
\affiliation{
    \institution{University of Sydney}
    \city{Sydney}
    \country{Australia}
}

\renewcommand{\shortauthors}{Jay Hwan Lee et al.}

\begin{abstract}
In emerging scientific computing environments, matrix computations of increasing size and complexity are increasingly becoming prevalent. However, contemporary matrix language implementations are insufficient in their support for efficient utilization of cloud computing resources, particularly on the user side. We thus developed an extension of the Julia high-performance computation language such that matrix computations are automatically parallelized in the cloud, where users are separated from directly interacting with complex explicitly-parallel computations. We implement lazy evaluation semantics combined with directed graphs to optimize matrix operations on the fly while dynamic simulation finds the optimal tile size and schedule for a given cluster of cloud nodes. A time model prediction of the cluster's performance capacity is constructed to enable simulations. Automatic configuration of communication and worker processes on the cloud networks allow for the framework to automatically scale up for clusters of heterogeneous nodes. Our framework's experimental evaluation comprises eleven benchmarks on an fourteen node (564~\makebox{vCPUs}) cluster in the AWS public cloud, revealing speedups of up to a factor of~\SI{5.1}{}, with an average~\SI{74.39}{\percent} of the upper bound for speedups.
\end{abstract}




\maketitle

\renewcommand\thefootnote{}
\footnotetext{\textbf{Abbreviations:} AWS, Amazon Web Service; HEFT, Heterogeneous Earliest Finish Time; CMM, Cloud Matrix Machine}

\renewcommand\thefootnote{\fnsymbol{footnote}}
\setcounter{footnote}{1}

\section{Introduction}\label{sec1}

Matrix multiplication is a fundamental operation in scientific
computing that has been extensively studied for
parallel computing~\cite{geijin1995, gallopoulos2016parallelism,
agarwal1994mult}.
Utilizing a commodity
cluster or the cloud for matrix computations is complex due to the
required concurrency control mechanisms.  In terms of performance, 
the bottleneck is often the network rather than the memory bandwidth of a single
node~\cite{dongarra2022evolution}. To maximize the performance of large
computations, splitting them into smaller operations leverages parallelism,
though at the cost of increased data communication. Striking a balance between
the data communication bottleneck of the network and multicore parallelism is a
key requirement to attain high performance in the cloud.
Ideally, programmers are shielded from such considerations,
if the underlying implementation of the sequential programming model can leverage the 
inherent parallelism for efficient execution in the cloud.
This particularly applies to matrix languages like MATLAB, Octave, and Julia, which
are designed for programmer productivity and
abstraction from the underlying multicore substrate, rather than
hand-optimization of hardware-dependent code.  To automatically
parallelize matrix computations, our approach uses lazy evaluation with hierarchical dependency analysis, online scheduling, and
simulation to effectively run tasks in parallel and reduce the impact of the
network on performance.  We implement this in Julia, a dynamic language for
high performance computing that is already equipped with a robust,
explicitly-parallel programming model~\cite{julia2017}.
The contributions of this paper are as follows.
\begin{itemize}
\item
The Julia cloud matrix machine~(Julia CMM) that extends the Julia
programming language with im\-plic\-itly-parallel matrix routines
for the cloud through an extension of the Julia matrix data type.
\item
An online simulation and run-time that employs a time model based on offline
profiling and regression analysis to predict task execution times. We
utilize this in conjunction with automated matrix parallelization and communicator configuration to both predict
the most efficient schedule of task parallelization across a cluster of cloud nodes
and to adjust the optimal distribution of communication and computation processes for each node.
\item
An extension of the HEFT scheduling algorithm that employs a node-level tile
cache and a dynamic tiling optimization to reduce the network-incurred
communication bottleneck.
\item
An extensive experimental evaluation on the AWS public cloud with homogeneous and heterogeneous clusters to show the validity of our approach.
\end{itemize}

\section{Related Work}

The computations in level~1 Basic Linear Algebra Subprograms (BLAS) involve vector-vector operations~\cite{level1BLAS}, whereas level~2 BLAS focuses on vector-processing machines~\cite{level2BLAS}. Level~3 BLAS is designed to leverage caches in a multi-memory hierarchy~\cite{level3BLAS}. To optimize cache utilization, matrices are divided into smaller matrices to achieve higher cache hit ratios~\cite{xianyi2012model,wang2013augem}. In Julia, OpenBLAS~\cite{OpenBLAS} is used as a multi-threaded level~3 BLAS implementation for linear algebra operations. Although research has been conducted on parallelizing BLAS operations for distributed systems~\cite{blackford1997scalapack,reddy2009heteropblas}, the existing methods provide low-level BLAS operations, a static communication model, and fixed tile sizes. The proposed Julia CMM, on the other hand, offers a high-level programming abstraction, a dynamic communication model, and adaptable tile sizes. Julia's standard library includes the \textsf{RemoteChannel} and \textsf{SharedArray} primitives for intra- and inter-node communication. However, programmers must manually parallelize their Julia code for cloud computing.
Several studies report that the network bandwidth for intra-node communication is the main limiting factor of matrix operations and try to model and optimize the communication cost~\cite{demmel2008ipdps,malik2016networkaware}. In addition to intra-node communication bandwidth, our Julia CMM considers the inter-node communication, dependencies of operations, and the number of communication and computation processes in optimization.

Numerous studies have explored automated parallelization of matrix operations, including optimization compilers~\cite{baskakov2021source,sato2011automatic} and generating optimal parallel algorithms for matrix-vector multiplication in neural networks~\cite{miyasaka2020vector}. Ongoing research focuses on optimizing matrix multiplication for Intel processors~\cite{nguyen2020Xeon,hemeida2020optimizing}, but this work does not involve implicit parallelism like that offered by Julia CMM. To address load imbalances, researchers have developed hybrid versions of breadth-first search (BFS) and depth-first search (DFS) algorithms for task-based parallelism~\cite{benson2014}, auto-tuning of sparse matrix-vector multiplication on multicore clusters with thread and process-based communication~\cite{shigang2015},
tiling of sparse matrix-matrix multiplication on GPUs~\cite{niu2022tilespgemm},
automation of stream parallelism in C++~\cite{streamParallelism}, and a modified selection prioritization HEFT algorithm for cloud environments~\cite{gupta2022efficient}. These frameworks, however, do not implement implicit parallelization for general matrix operations and may be more suitable for expert programmers than for domain experts or novice users.

Alternative implementations of the HEFT algorithm aim to address load imbalance issues in cloud networks, especially those that reduce throughput by creating idle resources when parent and child tasks with different input instances are processed in parallel~\cite{TPHEFT}. On the other hand, E-HEFT reduces the number of communication tasks, and its scheduler distributes tasks across nodes, followed by running a simulation~\cite{EHEFT}. To reduce communication tasks, we already utilize node-level tile cache and have implemented a separate scheduler and simulation as described in the paper. Conversely, EHEFT-R employs remapping resource allocation rules to synthesize optimal machines for ranked tasks~\cite{EHEFT-R}. However, ranking is not an effective optimization in a network where all machines are equal.

Scientific workflow research involves structuring computations connected through dependencies~\cite{taylor2007workflows} as a model for large-scale applications. Different tasks are scheduled to parallel resources, with task scheduling optimization being an NP-complete problem~\cite{ullman1975np}.
Heuristic scheduling algorithms are introduced for cloud computing service providers to minimize load imbalance between users of their large-scale cloud computing platforms~\cite{Kruekaew2020,zhan2012improved,PRIYA2019416}. While load balancing heuristics are applicable to Julia~CMM in improving the parallelizable performance, such heuristic algorithms aim to optimize the load balancing between all users of the cloud, while Julia~CMM's goal is to achieve the shortest makespan of matrix operations as a single user.
In a cloud environment, optimizing resource usage costs serves as an additional optimization point. Researchers proposed multi-objective scheduling algorithms~\cite{durillo2014multi,mahmoud2022multiobj} to meet multiple optimization objectives in a cloud environment, noting that doing so inevitably results in trade-offs.

\section{Julia CMM Framework}
\begin{figure*}[!t]
\centering
\includegraphics[width=\textwidth]{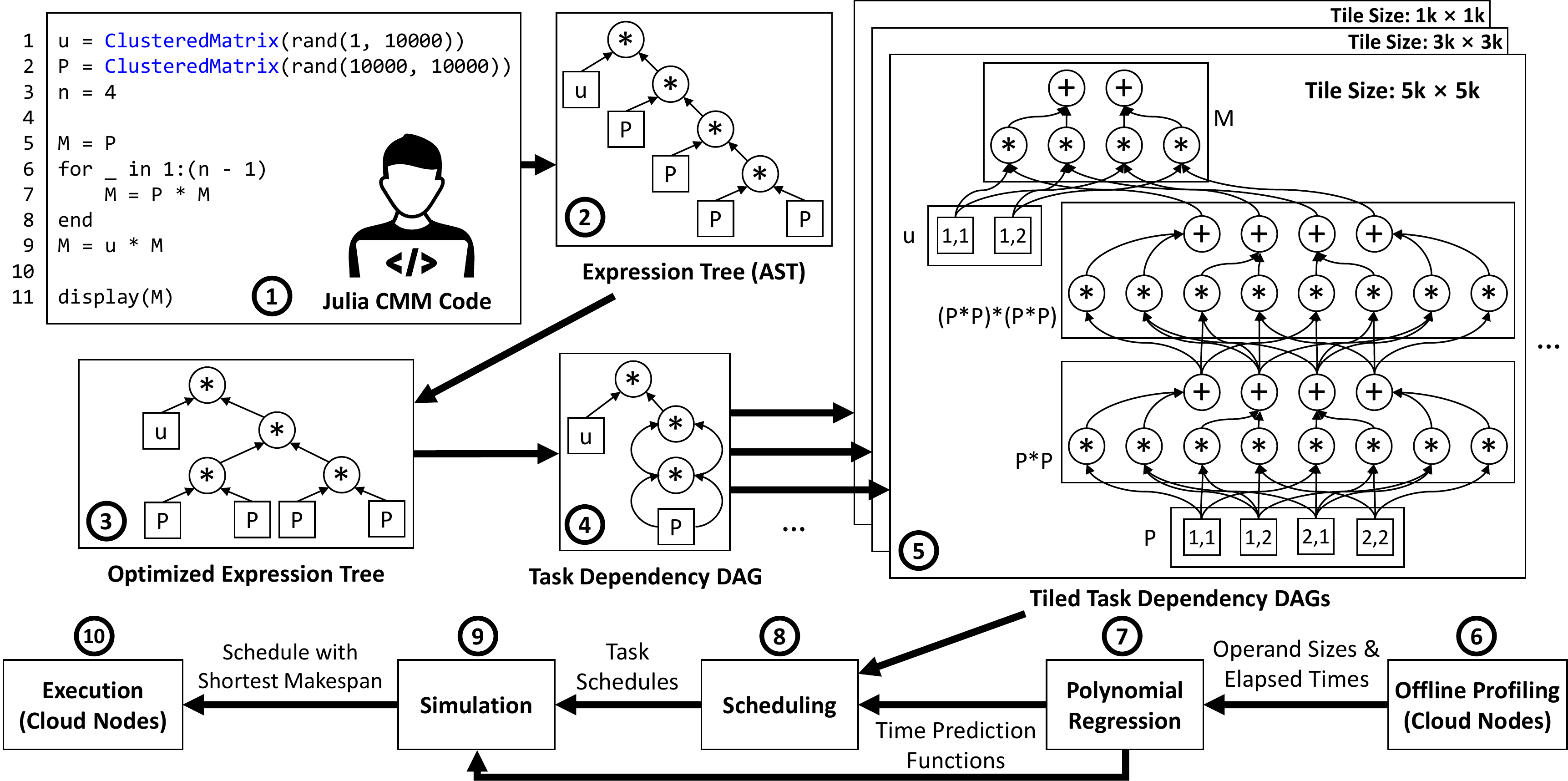}
\caption{Overview of Julia CMM, depicting the on-the-fly lowering of the Markov chain benchmark
source code~\CIRC{1} to
an expression tree~\CIRC{2} that is then optimized by employing exponentiation by squaring~\CIRC{3}.
Common subexpression elimination creates a task dependency DAG~\CIRC{4} which is lowered further
into tiled task dependency DAGs~\CIRC{5} for a series of standard tile sizes (here
\SIrange[range-units=repeat,range-phrase=$\,\times\,$,detect-weight=true]{1}{1}{\kilo{}},
\SIrange[range-units=repeat,range-phrase=$\,\times\,$,detect-weight=true]{3}{3}{\kilo{}}, and
\SIrange[range-units=repeat,range-phrase=$\,\times\,$,detect-weight=true]{5}{5}{\kilo{}}).
Online scheduling~\CIRC{8} and simulation~\CIRC{9} of each of the tiled task dependency DAGs is used
to determine the schedule with the smallest---predicted---makespan, which is then
executed on the cluster of cloud nodes~\CIRC{10}. The framework's time model 
is based on offline profiling~\CIRC{6} and regression analysis~\CIRC{7}.
}
\label{fig:overview}
\end{figure*}

In Step 1 of Figure~\ref{fig:overview} \STEP{1}, Julia CMM demonstrates minimal invasiveness for the user, as it only requires the casting of matrices to the provided \isrc{ClusteredMatrix} data type (lines 2-3) to parallelize sequential matrix code, such as the depicted Markov chain computation, without altering the actual computations (lines 5-9). The user writes code in a Jupyter notebook~\cite{Jupyter}, which serves as the frontend on their desktop or laptop. Julia CMM connects to the user's cloud instances via an MPI-like configuration file, listing the IP addresses of the cluster nodes and designating one node as the master.

Despite being initiated by users on their local machines, the actual computations of Julia CMM take place in the cloud. The framework adopts lazy evaluation, which means that during runtime, it substitutes the execution of matrix operations with the construction of an expression tree. The actual computation of operations is deferred until the result is needed. For example, in Figure~\ref{fig:overview} \STEP{1}, the \isrc{for} loop iterates three times, followed by the multiplication \isrc{u * M}, leading to the creation of the expression tree illustrated in Figure~\ref{fig:overview}
\STEP{2}. The \isrc{display} statement in line~11 triggers the actual computation of the tree, as it necessitates the output of the result to the user.

Lazy evaluation provides the advantage of aggregating operations before their execution, thereby broadening the {\em scope\/} of optimization and parallelization beyond individual operations, such as those in BLAS, to contiguous regions of operations in the execution trace. In the context of our example, this wider optimization scope allows the framework to identify the matrix exponentiation operation embedded in the expression tree (i.e., \isrc{P*P*P*P}) and rewrite it using exponentiation-by-squaring, as shown in \STEP{3} and discussed further in Section~\ref{sub:depGraph}. The elimination of the common subexpression \isrc{P*P} reduces the number of matrix multiplications in the resulting task dependency DAG, as illustrated in \STEP{4}.

The use of lazy evaluation in Julia~CMM widens its parallelization scope, allowing it to balance the amount of parallelism with the communication overhead involved in tiling matrix operands into sub-operands (herein called tiles). Tiling is necessary to reduce the granularity of work and utilize the parallel execution units of a cluster, but it increases communication overhead. Figure~\ref{fig:overview} \STEP{5} shows how Julia~CMM transforms a task dependency DAG into a series of tiled dependency DAGs, each encoding the {\em entire\/} task dependency DAG for a specific tile size. By creating tiled DAGs for a range of standard tile sizes, Julia~CMM conducts an online simulation to predict the makespan of each tiled DAG and selects the one with the shortest makespan for execution on the given cluster. This approach enables the framework to effectively balance parallelism and communication overhead in the trade-off.

To optimize the execution time of tiled DAGs, we use an online simulation with a time model that is obtained from offline profiling. This allows the re-use of profile data for known instance types and locations. We employ a modified version of the HEFT algorithm to generate task schedules that optimize the earliest finish times of tiled DAGs. After the master node determines the schedule with the shortest makespan, it uses worker processes on both the master and worker nodes to execute the schedule. The workers send their computed results back to the master node, from where the results are output to the user on their client machine.

\definecolor{lbcolor}{rgb}{0.9,0.9,0.9}
\begin{lstlisting}[basicstyle=\fontsize{7}{7}\ttfamily,
backgroundcolor=\color{lbcolor},
float,
numbers=left,
firstnumber=1,
numberfirstline=true,
caption={Implementation of the tile function}, 
captionpos=b,
label={lst:tile}]
function tile(P::ClusteredMatrix, tile_size::Tuple)
    mP, nP = size(P)
    mTile, nTile = tile_size
    mTiledP, nTiledP = cld(mP, mTile), cld(nP, nTile)
    tiledP = Matrix{ClusteredMatrix}(mTiledP, nTiledP)
    
    for j=1:nTiledP
        colStart = nTile * (j-1)+1
        colEnd = min(nTile * j, nP)
        colIdx = colStart:colEnd
        for i=1:mTiledP
            rowStart = mTile * (i-1)+1
            rowEnd = min(mTile * i, mP)
            rowIdx = rowStart:rowEnd
            tiledP[i,j] = view(P, rowIdx, colIdx)
        end
    end
    return tiledP
end
\end{lstlisting}

\subsection{Tree-rewriting, Tiling, Scheduling \& Simulation}\label{sub:depGraph}
In Figure~\ref{fig:overview}, we illustrate the process from \isrc{ClusteredMatrix}~in Julia~CMM to tiled dependency graphs on the master node using a Markov chain computation. The figure depicts references~\isrc{P} and~\isrc{u}, which point to \isrc{ClusteredMatrix} objects generated from random matrices based on the given input dimensions. As shown in \STEP{2} of Figure~\ref{fig:overview}, matrix~\isrc{P} is multiplied by itself three times before its final multiplication with matrix~\isrc{u}, represented in the expression tree.

The construction of an expression tree involves tracing the matrix operations performed at runtime by overloading the operators of the \isrc{ClusteredMatrix} data type. The input dependencies between the edges of the expression tree are used to build the task dependence graph. Julia CMM implements lazy evaluation, which defers the actual execution of matrix operations until a user-observable side-effect is reached or until the trace exceeds a predetermined threshold. Once this occurs, Julia~CMM pattern-matches the expression tree to rewrite sub-trees for which a more efficient equivalent is known to exist. This approach eliminates common subexpressions~\cite{dragonbook} and transforms series of multiplications into the exponentiation-by-squaring method~\cite{TAOCP:vol:2}. By contrast, standard Julia cannot perform this optimization unless the user manually replaces the \isrc{for} loop in Figure~\ref{fig:overview} with Julia's exponentiation operator.

In addition to exponentiation by squaring, further matrix algebraic optimizations in our implementation include
(1)~computing Moore-Penrose pseudo-inverse of matrices, particularly for use with the optimization of calculating non-square matrix computations for benchmarks with least square operations or sparse matrices~\cite{penrosePseudo};
(2) hierarchical singular value decomposition (HVSD) for optimizing dense matrix computations through reducing dimensionality while increasing parallelizability~\cite{HVSD};
(3)~computing matrix power through diagonalization, only for applicable
matrices that are capable of being diagonalized~\cite{diagonalization}; (4)~for
equations involving commutative matrix rings, employing an application of the
fast commutative matrix algorithm in reducing the number of multiplication
operations~\cite{commutativeFast}.
These algebraic optimizations are of high opportunity and profitability, as
shown by our experimental evaluation in Section~\ref{Evaluation}.

We note that as floating-point operations are not associative, there is the possibility of a loss of precision. As demonstrated in a study of Strassen's algorithm (which involves matrix partitioning) on Julia, double precision, \isrc{Float64}, is found to result in the most efficient precision particularly for larger matrix sizes, with a maximum absolute error of~\SI{e-14}{}for a matrix size of \SI{1}{\kilo{}}~\cite{precision}. We already utilize \isrc{Float64} in Julia~CMM to reduce the degree of error, but acknowledge it does not fully mitigate it.

The size of tiles used in partitioned matrix operations determines the
granularity and level of parallelism. Smaller tiles increase parallelism but
result in a larger number of suboperations and increased communication
overhead. We conducted experiments on both AWS cloud and our commodity cluster,
and observed that a tile size in the range of
\SIrange[range-units=single,range-phrase=--]{10}{50}{\percent} of the matrix
size provides optimal results for most benchmarks. However, tile sizes
exceeding~\SI[detect-weight=true]{50}{\percent{}} do not provide enough
parallelism. To determine the optimal tile size, we use time simulation with
time models instead of actual execution times, and generally use a sequence of
tile sizes (\SI[detect-weight=true]{1}{\kilo{}},
\SI[detect-weight=true]{3}{\kilo{}}, and~\SI[detect-weight=true]{5}{\kilo{}})
for a matrix size of~\SI[detect-weight=true]{10}{\kilo{}},
with~\SI[detect-weight=true]{5}{\kilo{}} being the best choice for most
benchmarks of that size.

In Figure~\ref{fig:overview}, the tile size of~\SI[detect-weight=true]{5}{\kilo{}} was predicted to have the shortest makespan for the input matrix of size~\SI[detect-weight=true]{10}{\kilo{}}. Using this tile size, the automatic tiling framework created four partitioned tiles of dimensions~\SIrange[range-units=repeat,range-phrase=$\,\times\,$,detect-weight=true]{5}{5}{\kilo{}} from matrix~\isrc{P}, and two tiles of dimensions~\SI{1}{}~$\times$~\SI[detect-weight=true]{5}{\kilo{}} from matrix~\isrc{u}. The dependency graph was updated to reflect the tiles, while maintaining the task dependencies.

To generate a tiled matrix, we use the algorithm shown in Listing~\ref{lst:tile}, which takes a matrix and tile size as arguments.
In lines~4--5, we make use of \isrc{cld}, which returns the smallest integer that is greater than or equal to the division of the input matrix's dimension by the input tile size's dimension, thereby resulting in the number of tiles to generate. 
A tiled matrix with tiles separated by tile dimensions (lines~8--15) is returned as a \isrc{ClusteredMatrix}, with tiles stored in a list following the input matrix's ordering. The process can be repeated for generating further sub-tiles.
	
\begin{figure}[hbt!]
\def\fwidth{73.2mm}
\def\lwidth{1.4mm}
\begin{subfigure}[t]{\columnwidth}
\centering%
\includegraphics[width=240pt]{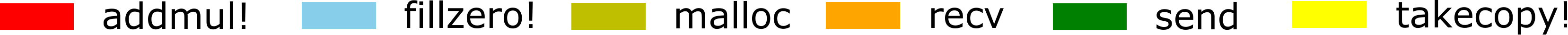}\vspace{-2mm}%
\end{subfigure}\vspace{1mm}
\centering
\begin{subfigure}[b]{0.45\textwidth}
\centering
\includegraphics[width=\fwidth]{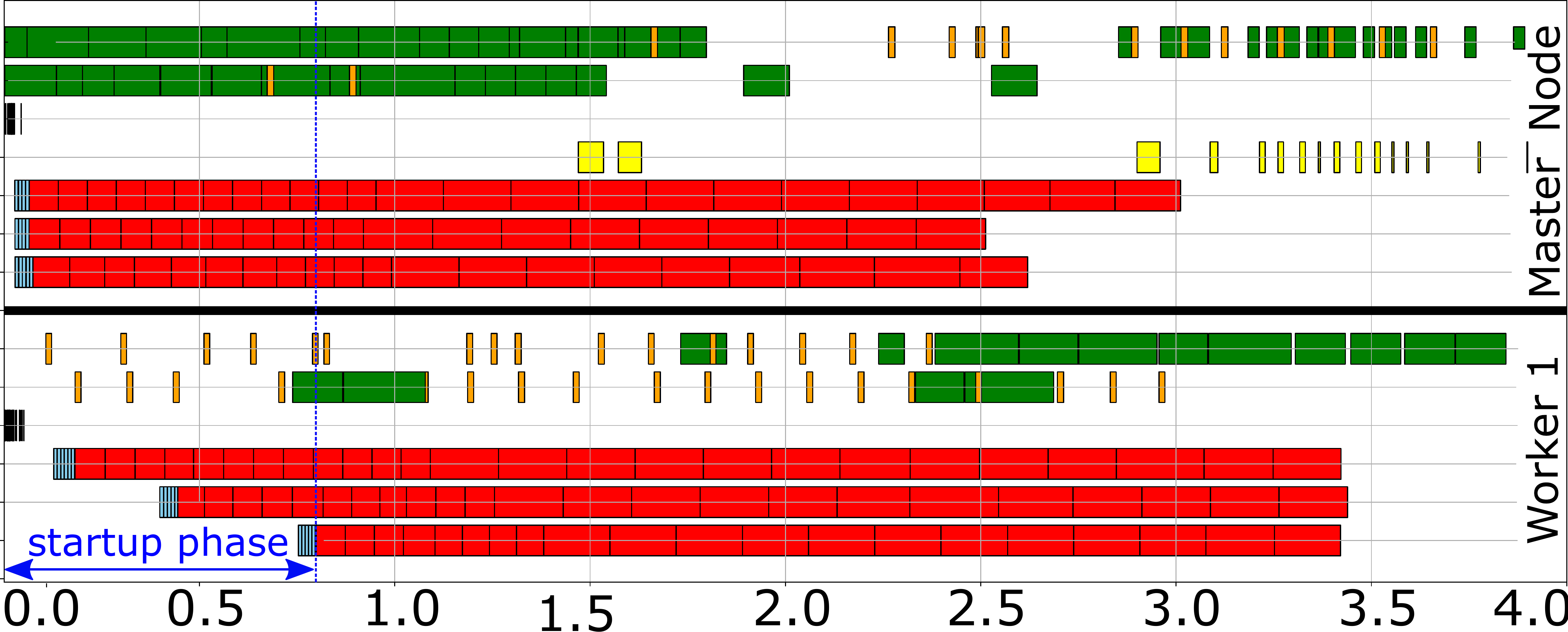}%
\caption{Two nodes, two communicators on the master node}
\label{fig:Markov2Node}
\end{subfigure}
\begin{subfigure}[b]{0.45\textwidth}
\centering
\includegraphics[width=\fwidth]{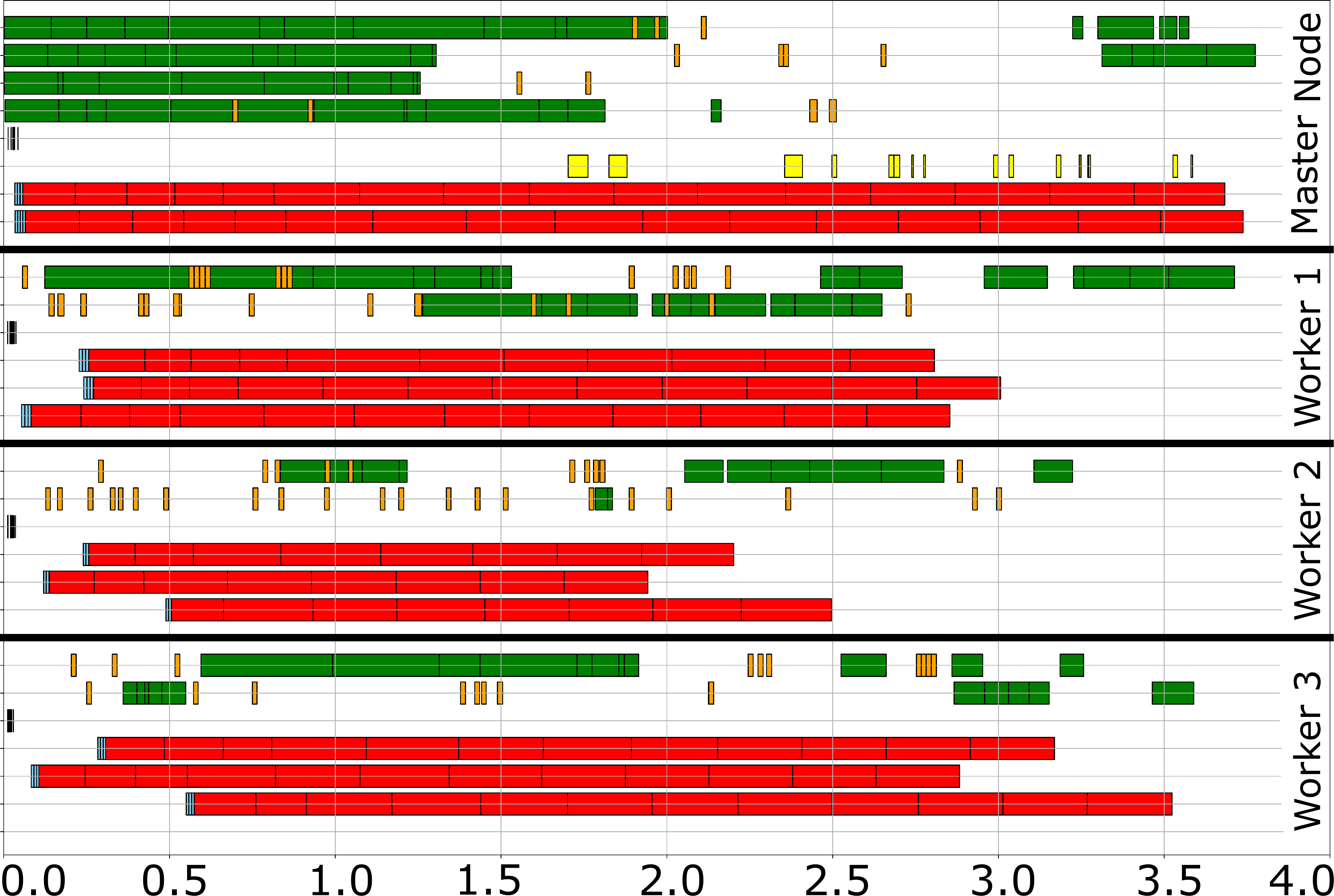}%
\caption{Four nodes, four communicators on the master node}
\label{fig:Markov4Node}
\end{subfigure}
\vspace{-2mm}
\caption{Markov chain benchmark task schedules for two cluster configurations
of two and four nodes on AWS c5.9xlarge instances using matrix
sizes of~\SI[detect-weight=true]{10}{\kilo{}} and tile
sizes of~\SI[detect-weight=true]{3}{\kilo{}}. Time flows from left to right.
Schedules are visualized on
a~\SI{4}{\second} timeline to facilitate comparisons.
Data send tasks (green) and matrix computations (red) dominate the schedules.
The length of receive tasks (orange) is shorter than the corresponding send tasks.
}
\label{fig:schedule}
\end{figure}

\subsection{Dynamic Tiling Optimization}\label{sub:dynTile}

To avoid underutilizing CPU cores on worker nodes, it is crucial to promptly propagate data and compute tasks to worker nodes as soon as the master node begins executing the selected schedule. We refer to the time duration from the start of a schedule until every worker has received computation tasks as the {\em startup phase}. As shown in Figure~\ref{fig:Markov2Node}, the upper half depicts the schedule of the master node, while the bottom half represents one worker node. Time progresses from left to right. The green rectangles represent data send operations, and the two green bars at the top of the figure indicate that the two communication processes in the master node continuously send data across the network to the worker node labeled "Worker~1". The red rectangles represent compute tasks, and the blue arrow in the bottom-left of the figure indicates that it takes almost \SI{0.75}{\s} into the schedule until all three worker processes of Worker~1 have begun computations. (Note that there are several matrix zeroing tasks represented by blue rectangles at the beginning of the schedule, but they are of small duration and are not the constraining factor for the makespan; the limiting factor is the time until data has arrived from the master node.)

The limited bandwidth of network communication poses a challenge in the efficient distribution of matrix data to worker nodes during the startup phase. In particular, large matrix tile sizes can monopolize the network and cause delays in communication with other nodes, resulting in idle workers or those that do not receive data at all. To address this issue, we propose a dynamic tiling approach in the startup phase. Our method involves the dynamic adjustment of tile sizes, where larger tiles are split into smaller sub-tiles to enable more operands to be sent across the network in the same amount of time. This generates a heterogeneous collection of tiles, with smaller tiles being prioritized over larger ones to maintain dependencies. During scheduling, sub-tiles are prioritized to bring work to all workers within a shorter duration. After all nodes receive work, sub-tiles are combined back into tiles in-place, allowing the original tile sizes to be utilized in subsequent matrix operations. This approach ensures that all workers are utilized efficiently during the startup phase, resulting in improved performance and reduced idle time.

Figure~\ref{fig:schedule} shows that during the startup phase, the worker nodes experience a significant downtime as they wait to receive data from the master node. In the absence of dynamic tiling, this downtime is even more severe and may result in some worker nodes not receiving any tasks, particularly in larger networks. However, by using dynamic tiling, smaller tasks are staged at the beginning of the schedule, reducing communication costs {\em per task} and minimizing the downtime. As described in Section~\ref{Evaluation}, the use of dynamic tiling leads to an overall speedup of $\ge$\SI[detect-weight=true]{10}{\percent} for most benchmarks.

We further extend the tiling process by parallelization. After a matrix is partitioned into tiles, each tile is concurrently split into sub-tiles until the desired tile sizes are achieved -- after which all sub-tiles are aggregated for use in the scheduling phase.

\subsection{Offline Profiling \& Time-Model Construction}\label{sub:profile}

Ensuring minimal total execution time in task allocation requires accurate prediction of each task's execution time. In this study, we employed off-line profiling on a selection of typical matrix operations to construct the Julia CMM time model. Time measurements were based on CPU time-stamp counters (TSCs). To overcome measurement errors across nodes where TSCs are typically unsynchronized, we employed a novel technique that ensures the measurement error remains below \SI{500}{\micro\second}~\cite{CloudprofilerArxiv}. The off-line profiling step was performed only once for a given cluster configuration.

To construct the time model, we utilized multivariate polynomial regression analysis based on the ordinary least squares (OLS) method~\cite{kreyszigAdvanced}. The amount of data used in this analysis is proportional to the matrix size and the number of floating point operations. Therefore, we made use of polynomial equations to predict the execution time. Table~\ref{tab:regression} outlines the classification for the interpolation equations used in constructing the time predictions, where the number of floating point operations is represented by a multivariate polynomial parameterized by the matrix size. The constants used in the equations were obtained through regression analysis, separately for each node and each node-to-node network link.
	
\begin{table}
\centering
\caption{Interpolation based on the operand and operator sizes. The first two columns are the left and right operands, followed by the operators and the interpolation equation.}
\label{tab:regression}
\vspace{-3mm}%
\footnotesize
\begin{tabular}{cccc}
\toprule
{\textbf{L Operand}} &
{\textbf{R Operand}} &
{\textbf{Operator}} &
{\textbf{Interpolation Eqn.}} \\
\cmidrule{1-4}
    $(n,1)$ & $(n,1)$ & $+,-,x$ & $a_0+a_{1}n$\\
    $(nm,1)$ & $(n,1)$ & $x$ & $a_0+a_{1}n+a_{2}m+a_{3}mn$\\
    $(m,n)$ & & $sin,cos$ & $a_0+a_{1}n+a_{2}m+a_{3}mn$\\
    $(m,n)$ & $1$ & $+,-,x,/$ & $a_0+a_{1}n+a_{2}m+a_{3}mn$\\
    $(m,n)$ & $(m,n)$ & $+,-,x$ & $a_0+a_{1}n+a_{2}m+a_{3}mn$\\
    $(m,n)$ & $(n,k)$ & $x$ & \multirow{2}{3cm}{\centering\textbf{$a_0+a_{1}n+a_{2}m+a_{3}k+...+a_{7}mnk$}} \\\\
\bottomrule
\end{tabular}
\end{table}

\subsection{Memory Management and Node-level Tile Cache\label{sub:nodeCache}}

When parallelizing a matrix operation, it is divided into smaller sub-operations that require certain operands, which may be needed by multiple succeeding operations (I.e., such an operand will have multiple successors in the dependency DAG, as illustrated in Figure~\ref{fig:overview}). This leads to redundant inter-node communication, where data is repeatedly sent to the same node. To address this issue, we introduced a node-level tile cache using a \isrc{SharedArray} (from the Julia standard library) in the main memory of each node. The \isrc{SharedArray} data type allows processes on the same node to share data, thus reducing the need for inter-node communication. The cache is controlled by the scheduler, with commands for placement and eviction of operands already part of the schedule, rather than guided by heuristics. The simulator accounts for the node-level tile caches in its predictions. Table~\ref{tab:nodecache} illustrates the performance improvements achieved by the node-level tile cache at various tile and network sizes.

\begin{table}
\centering
\footnotesize
\caption{
Simulated execution times of the Markov benchmark with and without
the node-level tile cache. Profiled data was obtained from
AWS c5.9x large instances (nodes) at a matrix size
of~\SI[detect-weight=true]{10}{\kilo{}}.
}
\label{tab:nodecache}
\renewcommand{\tabcolsep}{1mm}




\begin{tabular}{cc|rrrr|rrrr}

\toprule

\multirow{2}{*}{\rotatebox[origin=c]{40}{\textbf{Name}}} &
\multirow{2}{*}{\rotatebox[origin=c]{40}{\textbf{w/ cache}}} &
\multicolumn{4}{c}{\textbf{2 Nodes}} & 
\multicolumn{4}{c}{\textbf{8 Nodes}}\\

\cmidrule{3-10}

 &
 &
\SI{1}{\kilo{}} &
\SI{3}{\kilo{}} &
\SI{5}{\kilo{}} &
\SI{7}{\kilo{}} &
\SI{1}{\kilo{}} &
\SI{3}{\kilo{}} &
\SI{5}{\kilo{}} &
\SI{7}{\kilo{}} \\

 \midrule
 \multirow{3}{*}{\rotatebox[origin=c]{80}{\textbf{Markov}}} & Y (\si{\second}) & 4.92 & 3.61 & 2.97 & 12.69 & 3.77 & 2.92 & 1.57 & 10.23\\
  & & & & & & & & \\
 & N (\si{\second}) & 5.38 & 3.94 & 3.23 & 12.92 & 5.25 & 3.89 & 3.12 & 11.01\\

 \midrule
 \multirow{3}{*}{\rotatebox[origin=c]{80}{\textbf{Kmeans}}} & Y (\si{\second}) &8.85 &6.80 &6.10 &11.17 &5.74 &3.34 &2.37 &7.43 \\
   & & & & & & & & \\
 & N (\si{\second}) &9.25 &7.17 &6.78 &12.26 &6.21 &3.99 &2.51 &8.54 \\

 \midrule
 \multirow{3}{*}{\rotatebox[origin=c]{80}{\textbf{Hill}}} & Y (\si{\second}) &6.36 &3.85 &3.30 &11.80 &3.78 &2.81 &1.56 &9.92 \\
   & & & & & & & & \\
 & N (\si{\second}) &7.35 &4.25 &3.79 &12.13 &4.01 &3.05 &1.92 &10.72 \\

 \midrule
 \multirow{3}{*}{\rotatebox[origin=c]{80}{\textbf{Leontief}}} & Y (\si{\second}) &9.70 &8.78 &7.98 &15.81 &6.58 &5.05 &4.01 &11.77 \\
   & & & & & & & & \\
 & N (\si{\second}) &10.56 &9.82 &8.80 &16.62 &6.89 &5.49 &4.43 &12.60 \\

 \midrule
 \multirow{3}{*}{\rotatebox[origin=c]{80}{\textbf{Synth}}} & Y (\si{\second}) &6.47 &4.83 &4.82 &12.69 &3.97 &3.09 &1.40 &11.60 \\
   & & & & & & & & \\
 & N (\si{\second}) &7.23 &5.28 &5.62 &13.86 &5.14 &4.16 &3.34 &12.28 \\

 \midrule
 \multirow{3}{*}{\rotatebox[origin=c]{80}{\textbf{Reach.}}} & Y (\si{\second}) &8.69 &8.33 &7.24 &16.08 &5.94 &4.94 &4.19 &13.21 \\
   & & & & & & & & \\
 & N (\si{\second}) &9.11 &8.85 &7.42 &16.41 &6.38 &5.55 &4.56 &14.07 \\

 \midrule
 \multirow{3}{*}{\rotatebox[origin=c]{80}{\textbf{Hits}}} & Y (\si{\second}) &9.24 &6.99 &6.98 &16.69 &5.83 &4.27 &2.92 &10.28 \\
   & & & & & & & & \\
 & N (\si{\second}) &10.05 &7.68 &7.49 &17.08 &6.33 &5.09 &3.34 &11.07 \\

 \midrule
 \multirow{3}{*}{\rotatebox[origin=c]{80}{\textbf{BFS}}} & Y (\si{\second}) &27.31 &22.22 &18.55 &24.92 &14.88 &9.38 &6.70 &16.90 \\
   & & & & & & & & \\
 & N (\si{\second}) &29.44 &25.53 &20.43 &26.18 &16.93 &11.39 &8.06 &17.95 \\

 \midrule
 \multirow{3}{*}{\rotatebox[origin=c]{80}{\textbf{MM}}} & Y (\si{\second}) &36.91 &31.93 &23.47 &33.00 &13.90 &13.50 &10.77 &21.93 \\
   & & & & & & & & \\
 & N (\si{\second}) &39.04 &34.91 &26.53 &35.31 &16.45 &15.90 &12.90 &23.07 \\

 \midrule
 \multirow{3}{*}{\rotatebox[origin=c]{80}{\textbf{SPMV}}} & Y (\si{\second}) &43.05 &37.09 &27.14 &39.81 &25.64 &19.83 &14.78 &26.81 \\
   & & & & & & & & \\
 & N (\si{\second}) &48.24 &41.74 &34.47 &43.02 &30.58 &26.30 &28.42 &31.72 \\

 \midrule
 \multirow{3}{*}{\rotatebox[origin=c]{80}{\textbf{Montage}}} & Y (\si{\second}) &65.30 &53.14 &41.89 &73.28 &29.69 &24.21 &16.63 &63.58 \\
   & & & & & & & & \\
 & N (\si{\second}) &81.63 &71.50 &60.06 &85.80 &48.33 &47.85 &49.65 &77.81 \\
 
 \bottomrule
\end{tabular}\vspace{2mm}

\end{table}

\subsection{HEFT Scheduler Extension}\label{sub:HEFT}
To improve scheduling efficiency, we enhance the heterogeneous earliest-finish-time (HEFT~\cite{HEFT}) algorithm by incorporating the node-level tile cache awareness. The algorithm consists of two phases: task prioritization and assignment. During the prioritization phase, tasks are ranked recursively based on their average computation and communication costs across all nodes. In the assignment phase, tasks are assigned to processors such that the earliest finish time is selected. We employ the average internode communication costs and dynamically adjust the communication costs based on the presence of tiles in the node-level cache. In addition, we use the time predictions discussed in Section~\ref{sub:profile} to estimate the duration of each task. The scheduler accounts for communication time, which involves a send task from the sender and a receive task from the receiver, as shown in Figure~\ref{fig:schedule}. The algorithm gives priority to smaller tile sizes during task prioritization. The accuracy of the schedules generated is dependent on the accuracy of the cost predictions.

During the startup phase of the schedule, there is always a slight delay in which worker nodes must wait to receive data from the master. To minimize finish times, the scheduler prioritizes earliest finish time and initially favors worker processes on the master node, which has no communication requirements. However, since the capacity of the master node is limited, worker nodes are used to reduce finish times. The scheduler rejects scheduling on a node if the predicted combined communication and computation time is longer than waiting for a node where the tile is already present to become available. Because the master communicates with all workers, it has more dedicated communication processes to handle multiple communication tasks simultaneously, which reduces load imbalance caused by communication tasks waiting to be scheduled.

\begin{figure}
\def\fwidth{73.2mm}
\def\lwidth{1.4mm}
\begin{subfigure}[t]{\columnwidth}
\centering%
\includegraphics[width=240pt]{./schedule_diagrams/legend}\vspace{-2mm}%
\end{subfigure}\vspace{1mm}
\centering
\begin{subfigure}[b]{0.45\textwidth}
\centering
\includegraphics[width=\fwidth]{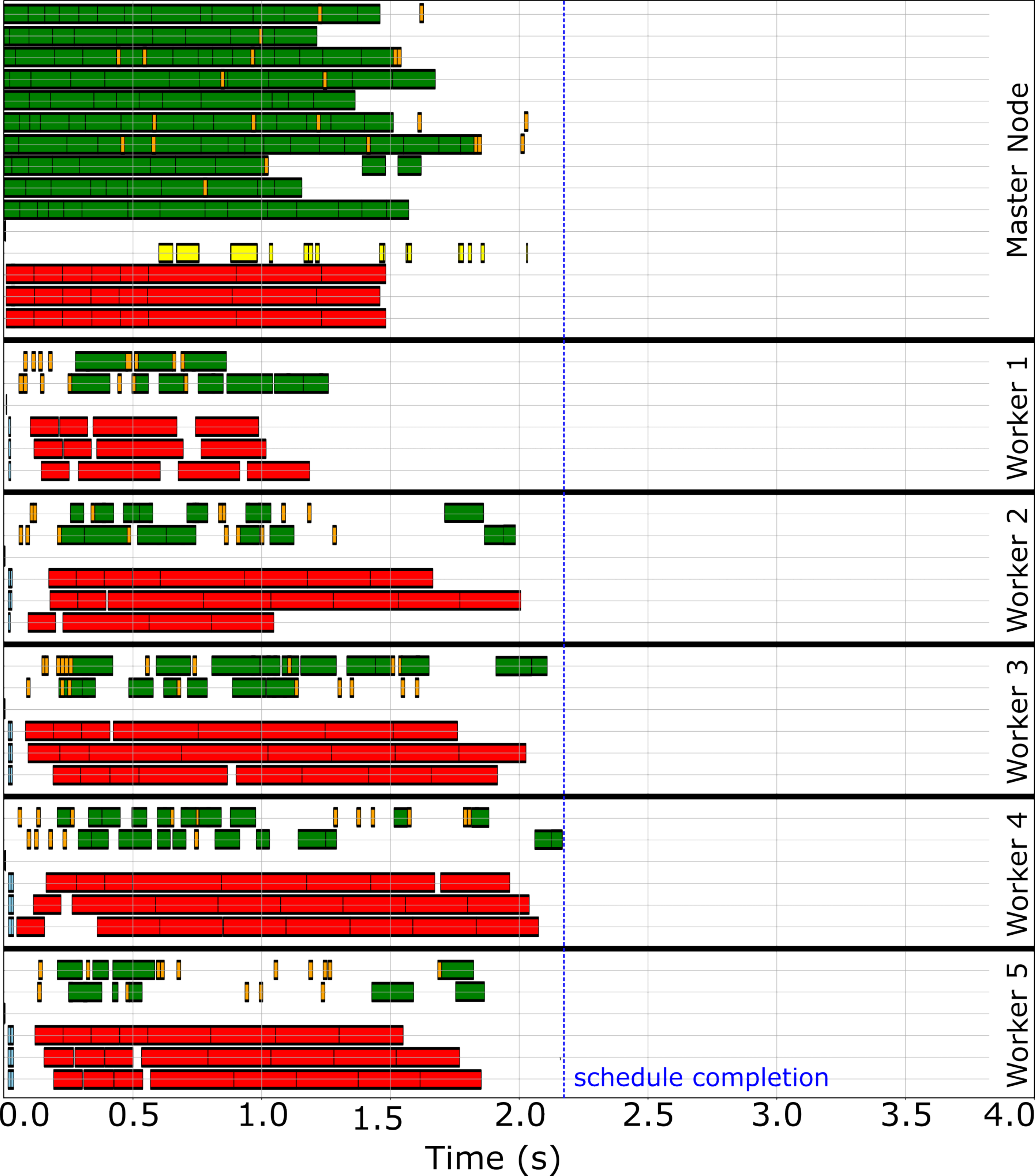}
\caption{Six nodes, cluster of heterogeneous nodes}
\label{fig:MarkovHetero}
\end{subfigure}
\begin{subfigure}[b]{0.45\textwidth}
\centering
\includegraphics[width=\fwidth]{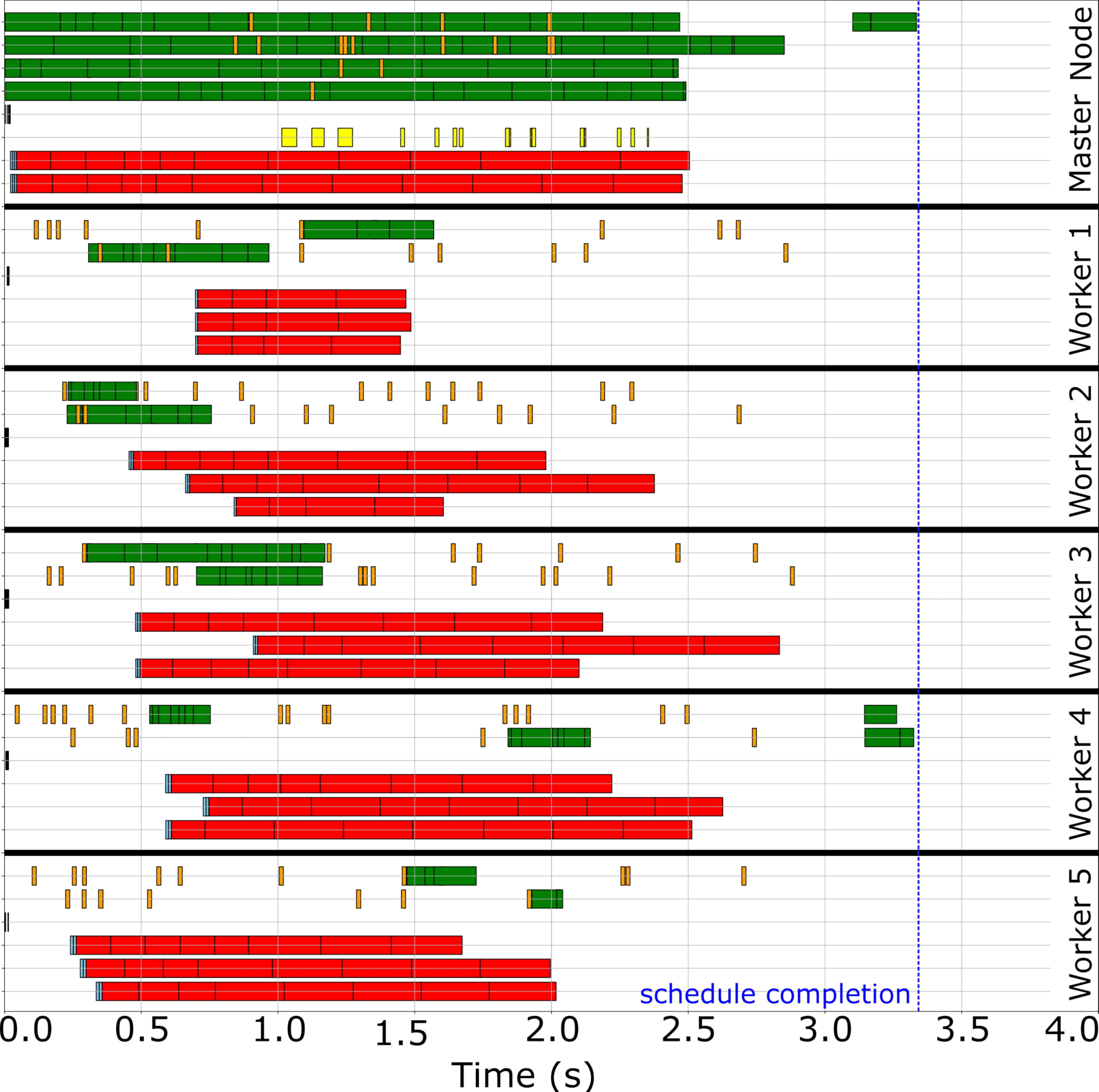}
\caption{Six nodes, cluster of homogeneous nodes}
\label{fig:Markov6Node}
\end{subfigure}
\vspace{-2mm}
\caption{Markov chain benchmark task schedules for a cluster of heterogeneous nodes and a cluster of homogeneous nodes. The heterogeneous cluster contained one AWS c5.24xlarge instance that was automatically selected by the framework as the master node, and all other (worker) nodes were AWS c5.9xlarge instances. The benchmark uses a matrix size of~\SI[detect-weight=true]{10}{\kilo{}} and tile
sizes of~\SI[detect-weight=true]{3}{\kilo{}}. Both schedules have six nodes in the network. Time flows from left to right.
Schedules are visualized on
a~\SI{4}{\second} timeline to facilitate comparisons. Julia~CMM's capacity for running on a heterogeneous cluster allows the deployment of a master node with a higher capacity for computation and communication, thereby shortening 
the startup phase from~\SI{0.81}{\second} to~\SI{0.22}{\second} resulting in a~\SI{359}{\percent} speedup, and the overall makespan from~\SI{3.4}{\second} to~\SI{2.2}{\second} resulting in a~\SI{155}{\percent} speedup.}
\label{fig:heteroSchedule}
\end{figure}

Figure~\ref{fig:MarkovHetero} shows the speedup improvement obtained from allocating more communication processes to the master node in a heterogeneous cluster, improving the schedule execution time from~\SI{3.4}{\second} to~\SI{2.2}{\second}. With the heterogeneous cluster, Julia~CMM makes use of an automatic communicator configuration (discussed further in Section~\ref{autoComm}) to automatically select the instance with the higher compute and network capacity as the master node.

For each node, the online simulator takes into account the duration of ongoing computation tasks, the number of available cores, the node's communication bandwidth, and whether the upcoming task's
dependencies are already cached in the node (if not, further communication
costs are incurred). If multiple candidate nodes are found where they have available cores and are ranked the same in bandwidth limit, the scheduler will tentatively tile the upcoming task and simulate the results.
The scheduler then selects the
node resulting in the earliest finish time, then moves on to the next
task.

We note that the scheduler generates different schedules {\em in parallel}, evaluating
different numbers of workers and communicators in the master node,
based on the upper bandwidth limit and number of cores determined during the profiling stage
(cf.~Table~\ref{tab:commWorker} and Table~\ref{tab:bandwidth}).
Once all schedules have been computed, the schedule that results in the
smallest makespan is selected.
This parallelization thereby minimizes the execution-time overhead of the
simulation. Our experimental evaluation in Section~\ref{Evaluation} shows that the
simulation overhead is below \SI{0.2}{\second} on average.

\subsection{Automatic Communicator Configuration}\label{autoComm}
In Julia~CMM, each node in the network is assigned a set number of communication and worker processes. One worker process comprises of the number of threads configured for BLAS, while one communication process contains one thread. Thus, a node with two communication processes and three worker processes has a total of 14~threads operating.
Our experiments make use of the Amazon EC2 service, where we utilize the instance type c5.9xlarge~\cite{AWS:c5}, each of which provides 36 vCPUs (18 physical cores) and a guaranteed~\SI{10}{\giga\bit\per\second} bandwidth in a shared network. Table~\ref{tab:gflops} shows the GFLOPS numbers for our framework based on the number of threads configured for an AWS c5.9xlarge instance. We find that resource oversubscription in a given node occurs at 16~threads or more. One c5.9xlarge instance has 36~vCPUs and 18~physical cores. Thus, adding more worker threads than the number of physical cores in a node causes thread contention utilizing resources, outweighing any potential benefit from introducing more worker processes.

As discovered from our dynamic tiling approach, the number of communication processes on the master node limits the extent to which parallelism can take place in a shared network. Thus, in scaling the framework to larger scales, the necessity of utilizing a cluster of heterogeneous nodes increases. In such a setup, it is advantageous for the framework to allocate the most powerful node as the master, as more communication processes can be allocated to each of its physical cores.
To facilitate this transition, we implement an automatic communicator detection and communication feature. During the profiling phase, for each node, Julia~CMM automatically obtains the maximum number of available threads based on the number of cores retrieved by Julia's \isrc{Sys.CPU_THREADS} command. A small benchmark is run in the network, where each node continuously sends and receives of empty communication requests. The number of communication processes for the master node is incrementally increased until either a saturated network bandwidth is detected (where further increases do not result in communication performance improvement), or until there are no available cores to allocate communication threads to (depending on the number of threads configured for BLAS, up to two worker threads are first allocated for the master node, and the remaining threads are checked for communication availability).
If an excess number of cores are left over after network bandwidth saturation, they are allocated to new worker processes.

Furthermore, as part of the communication profiling, it monitors the
inter-node network traffic flow and periodically records the maximum data
bandwidth between pairs of nodes,
the number of cores for each node, and the
number of tasks before saturation is reached. Scheduling utilizes such profiled
information to identify node-to-node links where a node's maximum bandwidth or number of cores is lower than its peers and rank the nodes in order of decreasing strength, prioritizing the placement of bandwidth capacity over computational power. 
The task and tile distribution is accordingly adjusted where lower ranked nodes are assigned fewer tasks
until either the bandwidth is saturated or the nodes are
unable to process more work, and favoring higher ranked nodes where bandwidth is not as big of a bottleneck.

Some degree of configuration by the user is required in the allocation of BLAS threads and maximum number of worker processes for the master and worker nodes. However, the automatic configuration opens up Julia~CMM to be pliable for clusters of heterogeneous nodes without necessitating deep manual intervention to match different node configurations.

\section{Experimental Evaluation}
\label{Evaluation}
Due to space constraints, for several aspects of our evaluation only a
subset of the data sets from the paper's accompanying technical
report~\cite{CMM:TR} is presented (and indicated as such). 

To evaluate Julia~CMM, we adopted benchmarks from a Cell
Octave benchmark set~\cite{burgstallerCellBench}, a set of highly
parallelizable applications. We specifically took the Markov, K-Means, Hill,
Leontief, Synth, Reachability, and Hits benchmarks and rewrote them in the
Julia language.
The Parboil Benchmark set~\cite{PARBOIL} contains a set of throughput
benchmarks for testing computer architecture in CUDA and OpenCL. We
specifically adapted the BFS, MM, and SPMV benchmarks, which use matrix
operations more predominantly than the others, and ported them to Julia.
Pegasus~\cite{pegasus1} is a programming language for workflow
applications such as CyberShake (earthquake
testing)~\cite{graves2011cybershake} and Sipht (searching untranslated
RNAs)~\cite{livny2008high}. Pegasus utilizes DAGs in reproducing and analyzing
scientific workflows, which share similarities with the schedules generated
with our Julia framework. One application that uses Pegasus is the Montage Image Mosaic Engine~\cite{MontageMosaic}, a tool for assembling astronomical input images into custom mosaics. 
We ported the Montage workflow written in Pegasus over to Julia that computes three separate color matrix channels to generate a small 2$\times$2 image mosaic.

We used the Amazon EC2 service with instance type c5.9xlarge~\cite{AWS:c5} as
our environment. This instance type was selected for having a
guaranteed~\SI{10}{\giga\bit\per\second} network bandwidth in a shared network,
and each instance provides 36~vCPUs (18~physical cores). We make use of a single instance type c5.24xlarge, which has 96~vCPUs and a guranteed~\SI{25}{\giga\bit\per\second} network bandwidth for our experiments with clusters of heterogeneous nodes. Julia~CMM
was developed on version~1.0 of the Julia language.

\begin{table}
\setlength{\tabcolsep}{1.0mm}
\centering
\footnotesize
\caption{GFLOPS performance of an AWS c5.9xlarge instance with Julia CMM, over 25 runs. We state the total number of threads, communicators per worker node, workers (and threads per worker), GFLOPS, and the coefficient of variation.}
\label{tab:gflops}
\vspace{-3mm}%

\begin{tabular}{lrrrrrr}\toprule
Total No.~threads &Comm. & Workers (threads) &GFLOPS Avg. &CoV (\SI{}{\percent})\\\midrule
4 & 1 & 3(1) & \SI{2257.8}{} &3.6\\
8 & 2 & 3(2) & \SI{1186.3}{} &5.9\\
12 & 3 & 3(3)& \SI{822.3}{} &5.3\\
14 & 2 & 3(4) & \SI{720.9}{} &7.3\\
16 & 4 & 3(4) & \SI{628.6}{} &4.1\\
32 & 4 & 7(4) & \SI{629.1}{} &6.5\\
64 & 8 & 7(8) & \SI{628.6}{} &6.8\\
\bottomrule
\end{tabular}
\end{table}

\begin{table*}[!htp]
\centering
\footnotesize
\caption{
Julia CMM performance for matrix sizes of
\SI[detect-weight=true]{10}{\kilo{}}~to~\SI[detect-weight=true]{30}{\kilo{}}, at a tile size of~\SI[detect-weight=true]{5}{\kilo{}}, for heterogeneous clusters of up to~14 nodes averaged over 20~runs.
Each cluster has one AWS c5.24xlarge instance as the master node, and AWS c5.9xlarge as the remaining nodes.
''Execution~Time'' depicts the execution times, ''Speedup'' depicts the relative speedup,
''Predicted~Time'' depicts the execution time predicted by the simulation, and ''Accuracy'' depicts the accuracy of the prediction.
The geometric mean of the relative speedups across all benchmarks is depicted at the bottom.
}
\label{tab:times}
\vspace{-3mm}
\renewcommand{\arraystretch}{0.30}
\setlength\tabcolsep{1.4pt}

\begin{tabular}{cc|ccccc|ccccc|ccccc|ccccc}

\toprule

\multirow{2}{*}{\textbf{Name}} &
\multirow{2}{*}{\textbf{Nodes}} &
\multicolumn{5}{|c}{\textbf{Execution~Time (\si{\second})}} &
\multicolumn{5}{|c}{\textbf{Speedup (\si{}{$\times$})}} &
\multicolumn{5}{|c}{\textbf{Predicted~Time (\si{\second})}} &
\multicolumn{5}{|c}{\textbf{Accuracy\ (\SI{}{\percent})}} \\

\cmidrule{3-22}

 &
 &
\SI{10}{\kilo{}} &
\SI{15}{\kilo{}} &
\SI{20}{\kilo{}} &
\SI{25}{\kilo{}} &
\SI{30}{\kilo{}} &
\SI{10}{\kilo{}} &
\SI{15}{\kilo{}} &
\SI{20}{\kilo{}} &
\SI{25}{\kilo{}} &
\SI{30}{\kilo{}} &
\SI{10}{\kilo{}} &
\SI{15}{\kilo{}} &
\SI{20}{\kilo{}} &
\SI{25}{\kilo{}} &
\SI{30}{\kilo{}} &
\SI{10}{\kilo{}} &
\SI{15}{\kilo{}} &
\SI{20}{\kilo{}} &
\SI{25}{\kilo{}} &
\SI{30}{\kilo{}} \\

\midrule
\multirow{8}{*}{\textbf{Markov}} & 1 &6.39 &12.52 &17.67 &20.23 &24.58 &1.00 &1.00 &1.00 &1.00 &1.00 &5.80 &11.91 &17.41 &19.86 &23.65 &110 &105 &102 &102 &104 \\
& 2 &5.41 &9.64 &13.84 &15.51 &17.03 &1.18 &1.30 &1.28 &1.30 &1.44 &4.62 &9.48 &13.37 &15.28 &16.88 &117 &102 &103 &102 &101 \\
& 4 &3.49 &7.16 &12.29 &14.25 &15.96 &1.83 &1.75 &1.44 &1.42 &1.54 &3.30 &6.25 &12.05 &13.83 &15.44 &106 &115 &102 &103 &103 \\
& 6 &2.59 &6.07 &11.42 &11.42 &13.33 &2.47 &2.06 &1.55 &1.77 &1.84 &2.27 &4.97 &10.41 &9.62 &10.99 &114 &122 &110 &119 &121 \\
& 8 &2.22 &6.87 &10.60 &10.49 &11.09 &2.88 &1.82 &1.67 &1.93 &2.22 &1.76 &6.70 &10.51 &8.96 &10.88 &126 &103 &101 &117 &102 \\
& 10 &1.82 &5.54 &9.19 &9.31 &9.22 &3.51 &2.26 &1.92 &2.17 &2.67 &1.55 &5.06 &8.25 &8.71 &8.33 &118 &110 &111 &107 &111 \\
& 12 &1.52 &5.09 &7.29 &7.74 &8.82 &4.22 &2.46 &2.43 &2.61 &2.79 &1.24 &4.31 &6.88 &6.60 &8.50 &122 &118 &106 &117 &104 \\
& 14 &1.28 &4.60 &6.93 &6.73 &7.89 &4.99 &2.72 &2.55 &3.01 &3.12 &1.07 &3.77 &6.76 &5.28 &7.46 &120 &122 &102 &127 &106 \\

\midrule
\multirow{8}{*}{\textbf{K-Means}} & 1 &10.09 &14.45 &18.95 &23.89 &26.14 &1.00 &1.00 &1.00 &1.00 &1.00 &9.34 &13.83 &18.20 &23.05 &25.56 &108 &105 &104 &104 &102 \\
& 2 &6.40 &10.31 &13.49 &18.98 &22.52 &1.58 &1.40 &1.41 &1.26 &1.16 &6.11 &9.66 &12.72 &18.43 &21.80 &105 &107 &106 &103 &103 \\
& 4 &5.18 &12.00 &13.65 &14.77 &16.27 &1.95 &1.20 &1.39 &1.62 &1.61 &5.14 &11.25 &13.30 &14.43 &15.75 &101 &107 &103 &102 &103 \\
& 6 &4.85 &8.21 &12.61 &14.56 &17.91 &2.08 &1.76 &1.50 &1.64 &1.46 &4.36 &7.51 &12.52 &14.53 &17.22 &111 &109 &101 &100 &104 \\
& 8 &2.94 &7.39 &11.00 &12.49 &15.80 &3.44 &1.96 &1.72 &1.91 &1.65 &2.79 &7.25 &10.25 &11.64 &15.74 &105 &102 &107 &107 &100 \\
& 10 &2.42 &8.14 &9.22 &11.07 &13.50 &4.18 &1.78 &2.05 &2.16 &1.94 &1.96 &7.88 &8.99 &10.15 &12.89 &123 &103 &103 &109 &105 \\
& 12 &2.20 &6.11 &9.91 &11.76 &12.39 &4.59 &2.37 &1.91 &2.03 &2.11 &2.07 &5.44 &9.61 &11.36 &12.37 &106 &112 &103 &104 &100 \\
& 14 &2.05 &4.77 &7.18 &10.74 &11.34 &4.92 &3.03 &2.64 &2.22 &2.31 &1.72 &3.93 &6.68 &10.51 &10.99 &119 &122 &107 &102 &103 \\

\midrule
\multirow{8}{*}{\textbf{Hill}} & 1 &7.48 &14.38 &19.43 &24.68 &29.18 &1.00 &1.00 &1.00 &1.00 &1.00 &7.03 &13.54 &19.20 &24.18 &28.34 &106 &106 &101 &102 &103 \\
& 2 &6.15 &9.07 &15.67 &15.59 &22.33 &1.22 &1.59 &1.24 &1.58 &1.31 &5.33 &8.48 &15.13 &15.52 &21.81 &116 &107 &104 &100 &102 \\
& 4 &3.60 &9.73 &14.26 &17.13 &18.78 &2.08 &1.48 &1.36 &1.44 &1.55 &3.16 &8.75 &14.15 &16.30 &17.84 &114 &111 &101 &105 &105 \\
& 6 &3.21 &5.64 &11.71 &14.26 &15.33 &2.33 &2.55 &1.66 &1.73 &1.90 &2.88 &4.75 &11.30 &13.75 &14.36 &112 &119 &104 &104 &107 \\
& 8 &2.73 &6.89 &11.19 &12.49 &14.82 &2.74 &2.09 &1.74 &1.98 &1.97 &2.61 &6.27 &10.66 &11.83 &14.09 &105 &110 &105 &106 &105 \\
& 10 &2.68 &4.50 &10.49 &10.56 &13.51 &2.80 &3.20 &1.85 &2.34 &2.16 &2.38 &3.77 &10.06 &9.90 &13.42 &112 &119 &104 &107 &101 \\
& 12 &2.19 &3.88 &8.72 &9.31 &13.16 &3.42 &3.70 &2.23 &2.65 &2.22 &1.74 &3.41 &8.47 &8.67 &12.89 &125 &114 &103 &107 &102 \\
& 14 &1.81 &3.19 &7.58 &8.76 &12.27 &4.13 &4.51 &2.56 &2.82 &2.38 &1.45 &2.70 &6.81 &7.84 &11.39 &125 &118 &111 &112 &108 \\

\midrule
\multirow{8}{*}{\textbf{Leontief}} & 1 &10.51 &17.31 &22.08 &26.66 &32.13 &1.00 &1.00 &1.00 &1.00 &1.00 &10.43 &16.48 &21.93 &26.65 &31.33 &101 &105 &101 &100 &103 \\
& 2 &9.34 &12.40 &18.27 &21.45 &23.98 &1.13 &1.40 &1.21 &1.24 &1.34 &8.88 &12.29 &18.05 &21.17 &23.15 &105 &101 &101 &101 &104 \\
& 4 &8.24 &11.81 &17.19 &17.47 &18.79 &1.28 &1.47 &1.28 &1.53 &1.71 &8.23 &10.88 &16.91 &17.42 &17.92 &100 &109 &102 &100 &105 \\
& 6 &7.47 &8.91 &15.03 &16.36 &16.75 &1.41 &1.94 &1.47 &1.63 &1.92 &7.07 &8.81 &14.50 &15.41 &15.87 &106 &101 &104 &106 &106 \\
& 8 &6.38 &8.27 &15.28 &16.52 &15.39 &1.65 &2.09 &1.45 &1.61 &2.09 &6.31 &7.83 &14.68 &16.38 &15.16 &101 &106 &104 &101 &101 \\
& 10 &5.58 &8.58 &14.07 &13.43 &14.35 &1.89 &2.02 &1.57 &1.99 &2.24 &5.09 &8.58 &13.93 &12.93 &14.18 &110 &100 &101 &104 &101 \\
& 12 &5.15 &6.00 &12.28 &14.95 &12.46 &2.04 &2.89 &1.80 &1.78 &2.58 &4.77 &5.39 &11.74 &14.40 &12.04 &108 &111 &105 &104 &104 \\
& 14 &4.66 &5.89 &12.40 &14.19 &11.56 &2.26 &2.94 &1.78 &1.88 &2.78 &4.24 &5.52 &11.53 &13.78 &11.12 &110 &107 &108 &103 &104 \\

\midrule
\multirow{8}{*}{\textbf{Synth}} & 1 &7.87 &12.38 &19.43 &27.53 &34.78 &1.00 &1.00 &1.00 &1.00 &1.00 &6.71 &12.32 &18.57 &27.23 &34.04 &117 &101 &105 &101 &102 \\
& 2 &5.86 &10.28 &13.71 &23.80 &24.45 &1.34 &1.20 &1.42 &1.16 &1.42 &5.06 &9.38 &12.90 &23.37 &24.13 &116 &110 &106 &102 &101 \\
& 4 &4.73 &9.16 &13.21 &19.26 &22.07 &1.66 &1.35 &1.47 &1.43 &1.58 &4.66 &8.40 &12.72 &19.17 &21.15 &102 &109 &104 &100 &104 \\
&6 &3.98 &6.56 &11.57 &17.63 &20.19 &1.98 &1.89 &1.68 &1.56 &1.72 &3.72 &5.66 &10.88 &17.61 &19.47 &107 &116 &106 &100 &104 \\
&8 &3.60 &5.38 &9.50 &14.72 &18.89 &2.19 &2.30 &2.04 &1.87 &1.84 &2.83 &4.82 &9.08 &14.30 &18.58 &127 &112 &105 &103 &102 \\
&10 &3.08 &4.96 &8.79 &13.54 &17.92 &2.55 &2.50 &2.21 &2.03 &1.94 &2.69 &4.00 &8.54 &13.21 &17.88 &115 &124 &103 &102 &100 \\
&12 &2.86 &4.36 &8.50 &10.74 &16.01 &2.75 &2.84 &2.28 &2.56 &2.17 &2.76 &3.93 &8.18 &10.32 &15.41 &104 &111 &104 &104 &104 \\
&14 &2.75 &4.14 &8.00 &9.78 &13.84 &2.86 &2.99 &2.43 &2.82 &2.51 &2.32 &4.07 &7.63 &9.01 &12.91 &118 &102 &105 &109 &107 \\

\midrule
\multirow{8}{*}{\textbf{Reachability}} & 1 &9.70 &16.78 &21.89 &24.96 &34.04 &1.00 &1.00 &1.00 &1.00 &1.00 &9.57 &16.76 &21.17 &24.16 &33.47 &101 &100 &103 &103 &102 \\
& 2 &8.44 &13.09 &18.50 &20.23 &27.01 &1.15 &1.28 &1.18 &1.23 &1.26 &8.36 &12.69 &17.67 &19.77 &26.33 &101 &103 &105 &102 &103 \\
& 4 &7.11 &12.27 &18.05 &18.67 &24.99 &1.36 &1.37 &1.21 &1.34 &1.36 &6.62 &11.30 &17.07 &17.73 &24.16 &107 &109 &106 &105 &103 \\
&6 &6.66 &12.18 &17.95 &17.67 &20.18 &1.46 &1.38 &1.22 &1.41 &1.69 &6.26 &11.84 &17.75 &16.81 &19.33 &106 &103 &101 &105 &104 \\
&8 &5.86 &10.17 &14.72 &16.81 &19.64 &1.66 &1.65 &1.49 &1.48 &1.73 &4.97 &9.93 &14.21 &16.66 &19.62 &118 &102 &104 &101 &100 \\
&10 &5.79 &9.72 &12.08 &14.20 &18.30 &1.68 &1.73 &1.81 &1.76 &1.86 &5.77 &9.22 &11.83 &13.91 &17.62 &100 &106 &102 &102 &104 \\
&12 &4.91 &9.42 &11.88 &13.07 &18.17 &1.98 &1.78 &1.84 &1.91 &1.87 &4.29 &9.38 &11.25 &12.88 &17.30 &115 &100 &106 &101 &105 \\
&14 &4.26 &7.96 &10.33 &11.53 &15.76 &2.28 &2.11 &2.12 &2.16 &2.16 &3.61 &6.99 &9.96 &11.34 &15.73 &118 &114 &104 &102 &100 \\

\midrule
\multirow{8}{*}{\textbf{Hits}} & 1 &9.86 &15.22 &21.94 &30.61 &38.95 &1.00 &1.00 &1.00 &1.00 &1.00 &8.90 &14.82 &21.81 &30.38 &38.26 &111 &103 &101 &101 &102 \\
& 2 &7.30 &13.84 &17.83 &25.68 &30.32 &1.35 &1.10 &1.23 &1.19 &1.28 &6.89 &13.04 &16.83 &25.66 &29.73 &106 &106 &106 &100 &102 \\
& 4 &6.56 &10.69 &17.63 &20.72 &25.76 &1.50 &1.42 &1.24 &1.48 &1.51 &5.57 &9.73 &17.17 &20.66 &25.57 &118 &110 &103 &100 &101 \\
&6 &6.47 &10.22 &16.69 &20.51 &22.09 &1.52 &1.49 &1.31 &1.49 &1.76 &5.84 &9.64 &15.85 &19.86 &21.74 &111 &106 &105 &103 &102 \\
&8 &5.42 &9.34 &16.27 &17.95 &21.80 &1.82 &1.63 &1.35 &1.71 &1.79 &4.68 &9.06 &15.36 &17.91 &21.06 &116 &103 &106 &100 &104 \\
&10 &4.37 &9.00 &11.86 &16.50 &19.25 &2.26 &1.69 &1.85 &1.86 &2.02 &3.83 &8.35 &11.53 &16.17 &18.29 &114 &108 &103 &102 &105 \\
&12 &4.32 &8.36 &11.27 &15.85 &18.28 &2.28 &1.82 &1.95 &1.93 &2.13 &4.14 &7.95 &10.63 &15.05 &17.77 &104 &105 &106 &105 &103 \\
&14 &4.23 &8.31 &10.51 &14.20 &18.50 &2.33 &1.83 &2.09 &2.16 &2.10 &3.78 &8.19 &9.76 &14.08 &17.83 &112 &102 &108 &101 &104 \\

\midrule
\multirow{8}{*}{\textbf{BFS}} & 1 &22.27 &31.92 &49.40 &73.04 &104.88 &1.00 &1.00 &1.00 &1.00 &1.00 &21.80 &31.20 &48.90 &72.89 &104.61 &102 &102 &101 &100 &100 \\
& 2 &20.14 &24.03 &36.26 &49.86 &71.27 &1.11 &1.33 &1.36 &1.46 &1.47 &19.95 &23.04 &35.92 &49.09 &70.36 &101 &104 &101 &102 &101 \\
& 4 &16.67 &20.19 &29.93 &38.50 &57.82 &1.34 &1.58 &1.65 &1.90 &1.81 &15.69 &20.13 &29.22 &37.51 &57.45 &106 &100 &102 &103 &101 \\
& 6 &14.44 &18.93 &22.47 &33.30 &47.13 &1.54 &1.69 &2.20 &2.19 &2.23 &13.77 &18.36 &22.33 &33.17 &46.21 &105 &103 &101 &100 &102 \\
& 8 &8.73 &14.33 &19.67 &27.23 &38.90 &2.55 &2.23 &2.51 &2.68 &2.70 &8.72 &13.95 &18.89 &27.10 &38.89 &100 &103 &104 &100 &100 \\
& 10 &8.69 &14.29 &19.79 &24.02 &32.69 &2.56 &2.23 &2.50 &3.04 &3.21 &7.90 &13.48 &18.90 &24.02 &32.52 &110 &106 &105 &100 &101 \\
& 12 &7.97 &13.35 &17.33 &23.76 &31.35 &2.79 &2.39 &2.85 &3.07 &3.34 &7.56 &12.94 &16.93 &23.36 &30.67 &105 &103 &102 &102 &102 \\
& 14 &7.16 &12.53 &16.95 &22.57 &29.57 &3.11 &2.55 &2.91 &3.24 &3.55 &6.34 &11.98 &16.52 &22.53 &28.70 &113 &105 &103 &100 &103 \\

\midrule
\multirow{8}{*}{\textbf{MM}} & 1 &34.56 &48.27 &70.58 &94.16 &125.38 &1.00 &1.00 &1.00 &1.00 &1.00 &33.85 &48.11 &69.88 &94.01 &125.18 &102 &100 &101 &100 &100 \\
& 2 &24.55 &32.86 &55.84 &74.10 &85.46 &1.41 &1.47 &1.26 &1.27 &1.47 &24.42 &32.52 &55.20 &73.24 &85.19 &101 &101 &101 &101 &100 \\
& 4 &21.63 &24.53 &41.68 &62.55 &73.12 &1.60 &1.97 &1.69 &1.51 &1.71 &21.32 &24.28 &40.96 &62.34 &72.53 &101 &101 &102 &100 &101 \\
& 6 &18.16 &17.50 &27.58 &53.00 &60.73 &1.90 &2.76 &2.56 &1.78 &2.06 &17.91 &16.95 &26.64 &52.48 &59.76 &101 &103 &104 &101 &102 \\
& 8 &13.15 &14.05 &21.72 &40.71 &49.48 &2.63 &3.43 &3.25 &2.31 &2.53 &12.95 &13.50 &21.35 &40.61 &48.89 &102 &104 &102 &100 &101 \\
& 10 &12.27 &13.53 &18.43 &31.20 &38.43 &2.82 &3.57 &3.83 &3.02 &3.26 &11.58 &13.04 &17.68 &30.64 &38.30 &106 &104 &104 &102 &100 \\
& 12 &11.32 &13.24 &16.76 &24.74 &33.91 &3.05 &3.65 &4.21 &3.81 &3.70 &10.83 &12.52 &16.28 &24.13 &33.65 &105 &106 &103 &103 &101 \\
& 14 &10.77 &12.70 &14.26 &19.57 &27.87 &3.21 &3.80 &4.95 &4.81 &4.50 &10.33 &12.18 &13.51 &19.16 &27.13 &104 &104 &106 &102 &103 \\

\midrule
\multirow{8}{*}{\textbf{SPMV}} & 1 &41.81 &50.39 &73.11 &125.32 &163.44 &1.00 &1.00 &1.00 &1.00 &1.00 &41.55 &49.51 &72.54 &124.57 &162.83 &101 &102 &101 &101 &100 \\
& 2 &28.51 &39.79 &54.22 &92.67 &124.43 &1.47 &1.27 &1.35 &1.35 &1.31 &28.14 &39.45 &54.06 &91.73 &123.69 &101 &101 &100 &101 &101 \\
& 4 &27.15 &32.47 &45.19 &73.98 &92.82 &1.54 &1.55 &1.62 &1.69 &1.76 &26.59 &31.72 &44.30 &73.52 &92.14 &102 &102 &102 &101 &101 \\
& 6 &22.06 &26.73 &40.95 &62.17 &80.02 &1.90 &1.89 &1.79 &2.02 &2.04 &21.65 &25.90 &40.11 &61.73 &79.74 &102 &103 &102 &101 &100 \\
& 8 &16.19 &19.19 &27.71 &52.71 &70.08 &2.58 &2.63 &2.64 &2.38 &2.33 &15.49 &18.21 &27.32 &51.97 &69.29 &105 &105 &101 &101 &101 \\
& 10 &14.79 &19.56 &27.15 &42.75 &59.79 &2.83 &2.58 &2.69 &2.93 &2.73 &14.67 &19.04 &26.26 &42.26 &58.95 &101 &103 &103 &101 &101 \\
& 12 &13.30 &18.36 &26.90 &39.63 &57.86 &3.14 &2.74 &2.72 &3.16 &2.82 &13.19 &18.17 &26.37 &39.32 &57.37 &101 &101 &102 &101 &101 \\
& 14 &12.49 &17.31 &23.56 &38.14 &55.44 &3.35 &2.91 &3.10 &3.29 &2.95 &11.74 &16.83 &23.46 &37.60 &55.29 &106 &103 &100 &101 &100 \\

\midrule
\multirow{8}{*}{\textbf{Montage}} & 1 &68.64 &79.16 &104.99 &160.32 &192.76 &1.00 &1.00 &1.00 &1.00 &1.00 &68.54 &78.90 &104.39 &159.73 &192.04 &100 &100 &101 &100 &100 \\
& 2 &43.08 &63.58 &77.11 &119.96 &131.98 &1.59 &1.25 &1.36 &1.34 &1.46 &42.42 &63.14 &76.98 &119.26 &131.54 &102 &101 &100 &101 &100 \\
& 4 &40.15 &49.70 &63.11 &107.95 &111.58 &1.71 &1.59 &1.66 &1.49 &1.73 &39.27 &48.79 &62.65 &106.96 &111.04 &102 &102 &101 &101 &100 \\
& 6 &32.33 &35.29 &49.91 &89.26 &89.22 &2.12 &2.24 &2.10 &1.80 &2.16 &32.27 &34.56 &49.38 &88.47 &89.00 &100 &102 &101 &101 &100 \\
& 8 &26.73 &25.86 &38.86 &67.46 &67.91 &2.57 &3.06 &2.70 &2.38 &2.84 &26.54 &25.05 &38.16 &67.24 &66.92 &101 &103 &102 &100 &101 \\
& 10 &20.57 &22.44 &32.33 &56.58 &59.27 &3.34 &3.53 &3.25 &2.83 &3.25 &20.22 &22.25 &31.62 &56.49 &58.64 &102 &101 &102 &100 &101 \\
& 12 &17.39 &19.54 &30.72 &51.96 &52.51 &3.95 &4.05 &3.42 &3.09 &3.67 &17.09 &19.42 &29.86 &51.66 &51.63 &102 &101 &103 &101 &102 \\
& 14 &16.13 &17.03 &26.39 &45.21 &47.96 &4.25 &4.65 &3.98 &3.55 &4.02 &15.19 &17.03 &25.80 &44.69 &47.55 &106 &100 &102 &101 &101 \\

\midrule
\multirow{8}{*}{\textbf{Average}} & 1 &-- &-- &-- &-- &-- &1.00 &1.00 &1.00 &1.00 &1.00 &-- &-- &-- &-- &-- &105 &103 &102 &101 &102 \\
& 2 &-- &-- &-- &-- &-- &1.31 &1.32 &1.30 &1.30 &1.35 &-- &-- &-- &-- &-- &106 &104 &103 &101 &102 \\
& 4 &-- &-- &-- &-- &-- &1.60 &1.51 &1.45 &1.52 &1.62 &-- &-- &-- &-- &-- &105 &107 &102 &102 &103 \\
& 6 &-- &-- &-- &-- &-- &1.85 &1.93 &1.69 &1.72 &1.88 &-- &-- &-- &-- &-- &107 &108 &103 &104 &105 \\
& 8 &-- &-- &-- &-- &-- &2.37 &2.20 &1.97 &1.99 &2.12 &-- &-- &-- &-- &-- &109 &105 &104 &103 &102 \\
& 10 &-- &-- &-- &-- &-- &2.68 &2.38 &2.24 &2.33 &2.42 &-- &-- &-- &-- &-- &110 &107 &104 &103 &103 \\
& 12 &-- &-- &-- &-- &-- &3.00 &2.70 &2.42 &2.53 &2.60 &-- &-- &-- &-- &-- &109 &107 &104 &104 &102 \\
& 14 &-- &-- &-- &-- &-- &3.29 &2.98 &2.71 &2.80 &2.85 &-- &-- &-- &-- &-- &114 &109 &105 &105 &103 \\

\bottomrule
\end{tabular}
\end{table*}
    
\begin{table*}
\centering
\footnotesize
\caption{Comparison of simulated execution times at different tile sizes for the Markov benchmark. 
The profiled data was obtained from eight AWS c5.9x large instances (nodes) at a matrix size
of~\SI[detect-weight=true]{10}{\kilo{}},
and the average over 20~runs per tile size is stated.
}
\label{tab:tilesizes}
\vspace{-3mm}
\renewcommand{\arraystretch}{1.0}


{ 
\setlength\tabcolsep{2pt}

\begin{tabular}{ccrrrrrrrrrrrrrrrrrrr}

\toprule

\multirow{2}{*}{\rotatebox[origin=c]{80}{\textbf{Name}}} &
\multirow{2}{*}{\rotatebox[origin=c]{80}{\textbf{Nodes}}} &
\multicolumn{17}{c}{\textbf{Tile size}} \\

\cmidrule{3-21}

 &
 &
500 &
\SI{1}{\kilo{}} &
\SI{1.5}{\kilo{}} &
\SI{2}{\kilo{}} &
\SI{2.5}{\kilo{}} &
\SI{3}{\kilo{}} &
\SI{3.5}{\kilo{}} &
\SI{4}{\kilo{}} &
\SI{4.5}{\kilo{}} &
\SI{5}{\kilo{}} &
\SI{5.5}{\kilo{}} &
\SI{6}{\kilo{}} &
\SI{6.5}{\kilo{}} &
\SI{7}{\kilo{}} &
\SI{7.5}{\kilo{}} &
\SI{8}{\kilo{}} &
\SI{8.5}{\kilo{}} &
\SI{9}{\kilo{}} &
\SI{9.5}{\kilo{}} \\

\midrule
\multirow{5}{*}{\rotatebox[origin=c]{80}{\textbf{Markov}}} & 1 &11.24 &10.20 &9.25 &8.86 &7.73 &7.71 &7.42 &6.95 &6.07 &5.03 &6.26 &8.14 &13.06 &20.76 &21.23 &21.60 &22.60 &22.95 &22.98 \\
 & 2 &7.75 &4.92 &4.74 &4.60 &4.42 &3.61 &3.26 &3.12 &3.28 &2.67 &4.13 &6.46 &8.65 &12.69 &13.58 &13.92 &15.19 &17.09 &17.17 \\
 & 4 &6.05 &4.14 &3.27 &2.94 &3.09 &2.81 &2.87 &2.42 &2.30 &1.96 &2.45 &3.21 &6.68 &11.35 &12.67 &13.48 &14.52 &16.39 &17.17 \\
 & 6 &6.22 &4.48 &3.49 &3.48 &3.33 &3.26 &2.78 &1.79 &1.75 &1.39 &1.62 &1.84 &8.40 &10.14 &11.75 &12.32 &12.86 &13.19 &13.67 \\
 & 8 &4.65 &3.77 &3.44 &3.10 &3.25 &2.92 &2.39 &1.39 &1.24 &1.27 &1.63 &1.23 &4.44 &10.23 &11.47 &11.98 &13.94 &14.42 &14.67 \\

\midrule
\multirow{5}{*}{\rotatebox[origin=c]{80}{\textbf{Kmeans}}} & 1 &14.13 &10.86 &11.25 &10.25 &10.32 &9.08 &7.41 &6.15 &5.77 &4.52 &7.15 &8.63 &12.87 &18.43 &19.81 &20.00 &20.95 &21.66 &22.34 \\
 & 2 &8.13 &6.85 &6.51 &6.46 &5.38 &3.80 &4.18 &4.33 &4.06 &4.10 &5.50 &5.73 &7.52 &11.17 &12.94 &13.72 &15.23 &16.04 &17.03 \\
 & 4 &8.10 &4.72 &5.20 &4.82 &4.10 &3.44 &3.92 &3.05 &3.55 &3.62 &4.49 &4.99 &5.75 &9.58 &10.52 &11.04 &11.48 &13.41 &14.32 \\
 & 6 &7.08 &4.84 &4.04 &3.60 &3.42 &3.63 &2.90 &3.05 &3.24 &2.51 &2.45 &2.59 &4.65 &7.68 &9.23 &10.21 &11.82 &12.07 &12.89 \\
 & 8 &4.55 &3.74 &3.02 &2.62 &2.68 &2.34 &1.85 &1.73 &1.73 &1.37 &1.65 &3.90 &5.10 &7.43 &7.84 &8.63 &8.72 &9.61 &10.20 \\

\midrule
\multirow{5}{*}{\rotatebox[origin=c]{80}{\textbf{Hill}}} & 1 &13.04 &10.17 &10.37 &9.95 &8.32 &7.90 &6.99 &6.36 &5.38 &4.76 &7.03 &8.82 &11.50 &17.86 &18.65 &19.33 &20.43 &21.34 &21.61 \\
 & 2 &7.71 &6.36 &5.40 &5.08 &4.47 &3.85 &3.84 &3.66 &2.93 &2.30 &3.69 &3.99 &9.22 &11.80 &12.81 &12.88 &13.15 &13.87 &14.20 \\
 & 4 &6.22 &5.35 &4.92 &3.55 &3.20 &3.44 &2.70 &2.54 &2.74 &2.10 &3.20 &4.49 &6.99 &9.42 &9.89 &10.23 &11.85 &12.67 &12.78 \\
 & 6 &5.73 &5.15 &4.14 &3.79 &3.45 &3.19 &2.80 &2.77 &2.20 &1.28 &2.50 &2.60 &7.92 &9.39 &9.43 &10.25 &11.08 &12.41 &12.60 \\
 & 8 &4.76 &3.78 &3.79 &3.66 &2.85 &2.81 &1.93 &1.61 &1.05 &0.86 &1.49 &4.70 &5.95 &9.92 &10.45 &10.71 &11.83 &12.71 &13.53 \\

\midrule
\multirow{5}{*}{\rotatebox[origin=c]{80}{\textbf{Leontief}}} & 1 &15.51 &13.88 &14.05 &13.09 &12.18 &11.90 &10.30 &9.43 &8.51 &7.20 &8.29 &11.70 &14.83 &21.83 &23.59 &24.30 &25.20 &26.60 &26.81 \\
 & 2 &11.34 &9.70 &9.57 &9.39 &8.17 &7.78 &8.15 &8.01 &7.50 &6.98 &8.20 &9.45 &13.47 &15.81 &16.66 &16.96 &18.92 &19.06 &19.39 \\
 & 4 &9.96 &7.19 &7.42 &7.22 &6.14 &5.76 &5.40 &5.25 &5.34 &5.32 &6.66 &7.56 &9.77 &13.86 &15.26 &16.00 &17.76 &18.78 &19.25 \\
 & 6 &9.28 &7.07 &6.41 &6.39 &6.39 &6.45 &5.68 &5.47 &4.81 &3.96 &3.74 &5.37 &11.40 &12.56 &12.63 &12.86 &13.12 &13.60 &14.58 \\
 & 8 &6.21 &6.58 &5.98 &5.77 &5.66 &5.05 &4.91 &4.23 &3.30 &3.01 &3.59 &4.74 &9.54 &11.77 &12.01 &12.28 &13.51 &14.94 &15.89 \\

\midrule
\multirow{5}{*}{\rotatebox[origin=c]{80}{\textbf{Synth}}} & 1 &13.60 &10.76 &10.61 &10.28 &9.95 &8.60 &8.05 &6.32 &4.40 &3.52 &6.63 &8.70 &12.65 &16.83 &17.67 &18.13 &19.16 &19.87 &20.30 \\
 & 2 &7.22 &6.47 &5.86 &5.43 &5.32 &4.73 &4.57 &3.74 &3.20 &3.82 &5.28 &3.98 &5.13 &12.69 &13.67 &14.39 &15.68 &16.72 &16.78 \\
 & 4 &7.42 &6.36 &4.65 &4.56 &4.04 &4.10 &4.05 &3.30 &3.17 &3.58 &3.96 &6.44 &6.31 &11.82 &13.66 &14.45 &15.22 &15.47 &15.51 \\
 & 6 &6.10 &4.33 &3.78 &3.53 &3.63 &3.02 &2.80 &1.96 &1.94 &1.22 &2.42 &3.52 &5.94 &10.44 &12.25 &12.85 &13.21 &14.65 &15.59 \\
 & 8 &3.65 &3.97 &3.12 &3.17 &2.83 &3.09 &2.52 &1.67 &1.21 &0.80 &1.44 &2.58 &5.43 &11.60 &12.41 &13.03 &14.54 &14.74 &15.13 \\

\midrule
\multirow{5}{*}{\rotatebox[origin=c]{80}{\textbf{Reachability}}} & 1 &14.97 &13.04 &13.31 &13.34 &12.11 &11.00 &10.50 &10.50 &9.32 &7.04 &8.77 &10.43 &15.92 &21.02 &21.10 &22.07 &22.54 &23.48 &24.03 \\
 & 2 &10.51 &8.69 &8.08 &7.54 &7.56 &8.33 &7.54 &7.67 &7.72 &6.24 &6.74 &7.22 &13.59 &16.08 &16.43 &16.80 &17.79 &18.46 &19.26 \\
 & 4 &9.56 &7.78 &7.07 &6.82 &7.13 &6.59 &6.45 &6.02 &6.07 &6.25 &6.86 &7.52 &10.25 &14.60 &15.05 &15.13 &16.24 &17.37 &17.65 \\
 & 6 &9.12 &7.24 &6.51 &6.19 &6.45 &6.00 &4.39 &4.60 &4.95 &4.83 &6.98 &5.21 &8.52 &13.99 &14.76 &14.92 &16.71 &17.98 &18.09 \\
 & 8 &6.63 &5.94 &6.14 &5.67 &4.98 &4.94 &3.76 &3.62 &3.82 &3.19 &5.55 &6.74 &10.36 &13.21 &14.98 &15.12 &15.65 &17.43 &17.71 \\

\midrule
\multirow{5}{*}{\rotatebox[origin=c]{80}{\textbf{Hits}}} & 1 &16.56 &14.85 &13.66 &13.32 &12.30 &11.70 &10.91 &10.37 &9.35 &8.70 &9.18 &9.73 &15.31 &21.49 &22.47 &22.71 &24.28 &24.89 &25.88 \\
 & 2 &11.15 &9.24 &8.44 &8.57 &7.12 &6.99 &7.61 &6.86 &6.46 &5.98 &6.26 &7.86 &10.22 &16.69 &18.13 &19.13 &19.84 &21.65 &22.54 \\
 & 4 &10.49 &8.51 &7.31 &7.32 &6.89 &7.78 &6.23 &6.21 &5.14 &4.94 &5.67 &7.34 &9.34 &14.92 &15.77 &16.62 &17.24 &18.42 &18.64 \\
 & 6 &9.53 &7.04 &6.72 &6.59 &6.09 &6.16 &4.62 &4.90 &4.80 &4.34 &4.77 &8.38 &9.76 &13.90 &14.95 &15.41 &15.62 &15.80 &15.92 \\
 & 8 &7.68 &5.83 &5.88 &5.32 &4.69 &3.27 &3.48 &3.01 &3.25 &2.62 &3.53 &3.09 &9.23 &10.28 &10.57 &11.13 &11.62 &12.73 &13.67 \\

\midrule
\multirow{5}{*}{\rotatebox[origin=c]{80}{\textbf{Parboil-BFS}}} & 1 &34.83 &33.30 &30.20 &29.15 &25.74 &23.42 &22.90 &22.54 &21.43 &20.46 &23.23 &25.67 &29.71 &30.38 &31.97 &34.45 &37.19 &37.28 &39.31 \\
 & 2 &29.16 &27.31 &23.26 &23.73 &23.99 &22.22 &21.80 &20.02 &19.62 &18.55 &19.16 &20.50 &22.04 &24.92 &27.84 &30.23 &30.93 &37.53 &38.60 \\
 & 4 &25.72 &21.10 &21.49 &19.99 &19.37 &18.17 &18.37 &16.25 &15.40 &14.09 &16.47 &19.88 &19.05 &24.15 &25.69 &24.35 &27.22 &29.26 &30.49 \\
 & 6 &23.38 &18.08 &18.85 &18.03 &17.21 &16.85 &16.22 &15.84 &12.57 &12.59 &14.29 &15.64 &17.32 &21.37 &22.31 &23.01 &23.37 &24.40 &27.76 \\
 & 8 &19.64 &15.88 &13.22 &14.66 &12.84 &11.38 &10.45 &10.34 &8.68 &6.30 &11.77 &11.32 &15.85 &16.90 &17.18 &18.69 &20.61 &22.04 &20.78 \\

\midrule
\multirow{5}{*}{\rotatebox[origin=c]{80}{\textbf{Parboil-MM}}} & 1 &45.98 &41.13 &39.46 &38.32 &38.49 &35.11 &34.27 &33.72 &30.98 &28.34 &35.34 &35.59 &36.98 &36.69 &39.48 &44.77 &45.62 &46.57 &47.13 \\
 & 2 &41.95 &36.91 &35.87 &34.36 &34.00 &31.93 &28.94 &28.13 &27.88 &23.47 &29.62 &30.43 &31.24 &33.00 &34.72 &37.47 &40.46 &45.33 &51.85 \\
 & 4 &36.48 &28.19 &27.22 &27.73 &27.11 &25.16 &25.23 &24.31 &22.20 &19.72 &23.27 &25.58 &27.86 &35.24 &35.08 &37.52 &37.90 &37.51 &42.41 \\
 & 6 &32.83 &28.29 &26.73 &27.46 &26.40 &25.06 &22.11 &21.13 &19.77 &17.30 &20.39 &23.12 &25.25 &28.95 &28.79 &27.60 &31.15 &31.09 &37.73 \\
 & 8 &28.64 &21.90 &21.68 &19.87 &19.24 &17.50 &16.97 &15.52 &13.85 &10.77 &15.23 &18.69 &20.84 &21.93 &22.81 &23.57 &25.91 &28.71 &28.52 \\

\midrule
\multirow{5}{*}{\rotatebox[origin=c]{80}{\textbf{Parboil-SPMV}}} & 1 &61.91 &52.10 &52.87 &49.14 &48.80 &47.02 &44.35 &42.15 &41.21 &36.69 &36.53 &40.58 &43.91 &47.94 &49.83 &49.17 &52.94 &55.01 &64.30 \\
 & 2 &49.33 &43.05 &42.35 &41.94 &40.81 &37.09 &35.96 &34.75 &31.46 &27.14 &29.70 &31.59 &32.78 &39.81 &40.21 &42.58 &48.82 &52.81 &58.16 \\
 & 4 &50.98 &43.59 &42.87 &41.85 &41.18 &39.27 &36.08 &32.24 &27.62 &25.85 &35.28 &35.70 &36.35 &37.80 &39.65 &40.34 &41.78 &44.19 &47.73 \\
 & 6 &47.59 &40.75 &37.56 &36.60 &34.49 &30.36 &30.71 &28.86 &24.18 &22.02 &34.58 &35.50 &34.71 &34.86 &36.50 &36.69 &43.29 &45.24 &48.75 \\
 & 8 &36.20 &34.64 &33.99 &33.34 &30.51 &29.83 &29.68 &22.42 &18.45 &14.78 &21.00 &21.34 &26.89 &26.81 &28.15 &33.70 &35.60 &39.73 &42.59 \\

\midrule
\multirow{5}{*}{\rotatebox[origin=c]{80}{\textbf{Montage}}} & 1 &96.15 &81.31 &85.11 &84.38 &80.53 &75.44 &69.87 &66.05 &63.74 &58.50 &63.84 &78.31 &78.83 &80.60 &86.67 &88.37 &86.77 &95.35 &95.82 \\
 & 2 &88.53 &75.30 &73.81 &71.66 &68.28 &63.14 &62.03 &59.44 &48.86 &41.89 &58.88 &61.01 &66.05 &73.28 &72.28 &75.77 &79.60 &80.29 &87.35 \\
 & 4 &90.48 &85.39 &79.02 &74.91 &69.35 &68.05 &63.92 &52.45 &45.62 &37.89 &61.34 &68.35 &69.14 &70.98 &71.08 &70.35 &73.60 &73.36 &81.59 \\
 & 6 &81.23 &74.62 &73.44 &63.17 &51.69 &48.60 &37.84 &32.70 &28.87 &22.20 &42.51 &54.33 &61.02 &63.16 &66.30 &71.94 &75.25 &77.37 &81.55 \\
 & 8 &77.83 &64.69 &60.62 &54.24 &47.42 &42.21 &37.21 &31.79 &22.77 &16.63 &39.39 &56.47 &61.48 &63.58 &68.17 &65.29 &69.84 &73.94 &76.09 \\

\bottomrule

\end{tabular}
}

\end{table*}

\begin{figure*}[hbt!]
\includegraphics[width=500pt]{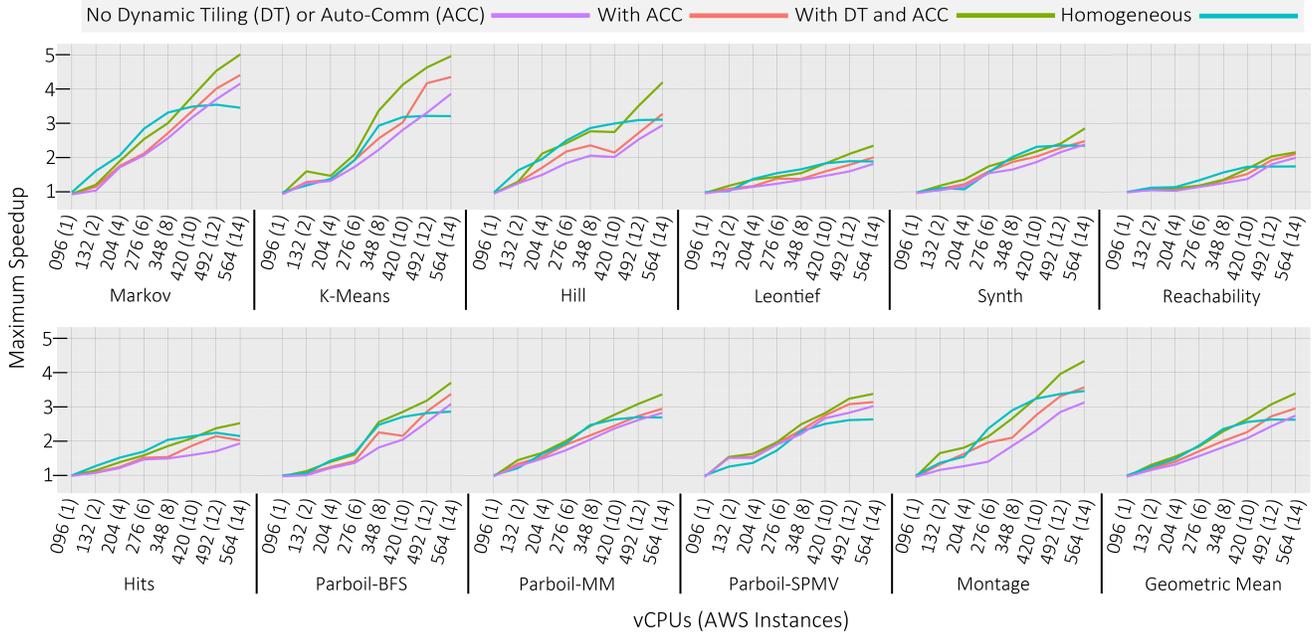}
\vspace{-3mm}%
\caption{Julia~CMM speedups without any optimizations, with the automatic communicator configuration (ACC), and with both ACC and dynamic tiling, on a heterogeneous network of
one c5.24xlarge AWS instance node (96~vCPUs) to up to fourteen nodes with c5.9xlarge AWS instances (564~vCPUs total).
Julia~CMM (the entire computation, including online simulation) is
compared against the time to execute the benchmark in vanilla Julia on a single node.
The homogeneous line shows the time to execute on a homogeneous network of c5.9xlarge AWS instances. Its x-axis values differ by the number of vCPUs, in that they go from 036(1)~to~288(8) in intervals of 36.
All benchmarks were run with a matrix size
of~\SI[detect-weight=true]{10}{\kilo{}}, with simulated tile sizes ranging
from~\SI[detect-weight=true]{1}{\kilo{}}
to~\SI[detect-weight=true]{5}{\kilo{}}. The tile size that resulted in the
highest performance in the online simulation is automatically selected by the
framework. The last diagram depicts the geometric mean 
across all benchmarks.}
\label{fig:speedup}
\end{figure*}

Table~\ref{tab:times} demonstrates the accuracy of the simulation is comparable with the execution of the benchmarks, with predicted values usually comprising a~\SIrange[range-units=single,range-phrase=--]{5}{20}{\percent} difference from actual values. The simulation thereby is at a point where its time-model prediction is useful in effective task scheduling,
particularly with the average simulation time being less than~\SI{0.2}{\second} acting as a marginal overhead. There is overhead in managing the task dependencies during execution, causing the node-level tile cache to be slightly underestimated in the simulation, contributing to the slight accuracy deviation.

\begin{table}[b]
\centering
\footnotesize
\caption{Performance improvement of the dynamic tiling optimization
for matrix sizes of~\SI[detect-weight=true]{10}{\kilo{}},~\SI[detect-weight=true]{15}{\kilo{}}, and~\SI[detect-weight=true]{20}{\kilo{}}, at a tile size of~\SI[detect-weight=true]{5}{\kilo{}}. The columns depict the original simulated execution time, the improved one with dynamic tiling alone and no other optimizations, and the relative speedup compared to without the optimization. The geometric mean of the speedups for all 11 benchmarks are listed at the end. 
}
\label{tab:dynamic}
\vspace{-3mm}
\renewcommand{\arraystretch}{0.85}



{ 
\setlength\tabcolsep{2pt}

\begin{tabular}{cc|ccc|ccc|rrr}

\toprule

\multirow{2}{*}{\rotatebox[origin=c]{80}{\textbf{Name}}} &
\multirow{2}{*}{\rotatebox[origin=c]{80}{\textbf{Nodes}}} &
\multicolumn{3}{c}{\textbf{Original (\si{\second})}} &
\multicolumn{3}{|c}{\textbf{Optimized (\si{\second})}} &
\multicolumn{3}{|c}{\textbf{Speedup (\si{}{$\times$})}} \\

\cmidrule{3-11}

 &
 &
\SI{10}{\kilo{}} &
\SI{15}{\kilo{}} &
\SI{20}{\kilo{}} &
\SI{10}{\kilo{}} &
\SI{15}{\kilo{}} &
\SI{20}{\kilo{}} &
\SI{10}{\kilo{}} &
\SI{15}{\kilo{}} &
\SI{20}{\kilo{}} \\

\midrule
\multirow{5}{*}{\rotatebox[origin=c]{80}{\textbf{Markov}}} &  1 &6.15 &11.99 &17.38 &6.04 &11.91 &16.81 &1.02 &1.01 &1.03 \\
 & 2 &4.80 &10.49 &15.97 &2.97 &9.10 &14.46 &1.61 &1.15 &1.10 \\
 & 4 &4.17 &8.74 &14.54 &2.86 &7.79 &13.09 &1.46 &1.12 &1.11 \\
 & 6 &3.00 &7.63 &14.24 &1.79 &6.43 &12.64 &1.67 &1.19 &1.13 \\
 & 8 &2.60 &7.83 &13.34 &1.57 &6.45 &11.66 &1.66 &1.21 &1.14 \\       

\midrule
\multirow{5}{*}{\rotatebox[origin=c]{80}{\textbf{Kmeans}}} &  1 &7.64 &13.62 &18.23 &7.55 &13.18 &18.23 &1.01 &1.03 &1.00 \\
 & 2 &6.78 &10.92 &16.92 &6.10 &9.49 &15.06 &1.11 &1.15 &1.12 \\
 & 4 &5.58 &10.09 &16.27 &5.12 &9.17 &14.04 &1.09 &1.10 &1.16 \\
 & 6 &3.86 &9.74 &15.64 &3.51 &8.42 &13.78 &1.10 &1.16 &1.14 \\
 & 8 &2.51 &7.70 &13.81 &2.37 &6.80 &11.63 &1.06 &1.13 &1.19 \\     

\midrule
\multirow{5}{*}{\rotatebox[origin=c]{80}{\textbf{Hill}}} &  1 &6.39 &12.25 &17.59 &6.12 &11.85 &17.43 &1.04 &1.03 &1.01 \\
 & 2 &3.79 &8.98 &15.02 &3.30 &8.10 &13.80 &1.15 &1.11 &1.09 \\
 & 4 &3.28 &8.98 &14.81 &2.80 &7.92 &13.42 &1.17 &1.13 &1.10 \\
 & 6 &2.06 &7.66 &13.99 &1.78 &6.63 &12.02 &1.16 &1.15 &1.16 \\
 & 8 &1.92 &7.82 &13.23 &1.56 &6.69 &11.73 &1.23 &1.17 &1.13 \\

\midrule
\multirow{5}{*}{\rotatebox[origin=c]{80}{\textbf{Leontief}}} &  11 &10.06 &15.58 &21.22 &9.50 &15.09 &20.63 &1.06 &1.03 &1.03 \\
 & 2 &8.80 &14.23 &20.04 &7.98 &12.14 &18.49 &1.10 &1.17 &1.08 \\
 & 4 &6.73 &12.27 &18.35 &6.32 &10.31 &15.37 &1.06 &1.19 &1.19 \\
 & 6 &5.45 &11.24 &16.74 &4.96 &9.58 &14.61 &1.10 &1.17 &1.15 \\
 & 8 &4.43 &9.50 &16.43 &4.01 &7.90 &14.71 &1.11 &1.20 &1.12 \\   

\midrule
\multirow{5}{*}{\rotatebox[origin=c]{80}{\textbf{Synth}}} &  1 &6.08 &11.92 &17.58 &5.84 &11.76 &17.26 &1.04 &1.01 &1.02 \\
 & 2 &5.62 &9.71 &16.09 &4.82 &8.33 &14.86 &1.17 &1.17 &1.08 \\
 & 4 &6.16 &10.07 &15.17 &4.58 &9.03 &13.52 &1.34 &1.12 &1.12 \\
 & 6 &4.76 &7.98 &14.08 &2.22 &6.88 &12.52 &2.15 &1.16 &1.12 \\
 & 8 &3.34 &6.56 &12.57 &1.40 &5.49 &11.08 &2.38 &1.19 &1.13 \\         

\midrule
\multirow{5}{*}{\rotatebox[origin=c]{80}{\scriptsize\textbf{Reachability}}} &  1 &9.36 &15.33 &20.66 &9.02 &14.82 &19.96 &1.04 &1.03 &1.03 \\
 & 2 &7.42 &13.17 &19.00 &7.24 &11.57 &17.22 &1.02 &1.14 &1.10 \\
 & 4 &7.65 &12.98 &18.50 &7.25 &11.48 &17.13 &1.05 &1.13 &1.08 \\
 & 6 &7.05 &11.68 &17.09 &6.83 &10.72 &15.00 &1.03 &1.09 &1.14 \\
 & 8 &4.56 &9.95 &15.85 &4.19 &8.43 &13.24 &1.09 &1.18 &1.20 \\ 

\midrule
\multirow{5}{*}{\rotatebox[origin=c]{80}{\textbf{Hits}}} &  1 &9.87 &15.09 &21.16 &9.72 &14.76 &20.83 &1.02 &1.02 &1.02 \\
 & 2 &7.49 &12.93 &19.09 &6.98 &11.44 &17.60 &1.07 &1.13 &1.08 \\
 & 4 &6.24 &11.72 &17.62 &5.94 &10.13 &16.27 &1.05 &1.16 &1.08 \\
 & 6 &5.67 &10.84 &16.92 &5.34 &9.02 &14.98 &1.06 &1.20 &1.13 \\
 & 8 &3.34 &9.75 &15.74 &2.92 &7.93 &13.61 &1.15 &1.23 &1.16 \\ 

\midrule
\multirow{5}{*}{\rotatebox[origin=c]{80}{\textbf{BFS}}} &  1 &22.06 &31.79 &48.52 &21.48 &30.92 &47.3 &1.03 &1.03 &1.03 \\
 & 2 &20.43 &27.19 &45.07 &18.55 &23.49 &35.44 &1.10 &1.16 &1.27 \\
 & 4 &17.08 &22.69 &39.69 &14.09 &18.96 &28.14 &1.21 &1.20 &1.41 \\
 & 6 &15.54 &19.15 &37.41 &12.59 &16.97 &21.84 &1.23 &1.13 &1.71 \\
 & 8 &8.06 &17.29 &36.87 &6.70 &12.59 &17.74 &1.20 &1.37 &2.08 \\ 

\midrule
\multirow{5}{*}{\rotatebox[origin=c]{80}{\textbf{MM}}} &  1 &33.12 &47.51 &71.62 &31.92 &46.56 &69.82 &1.04 &1.02 &1.03 \\
 & 2 &26.53 &35.26 &62.97 &23.47 &32.05 &52.92 &1.13 &1.10 &1.19 \\
 & 4 &23.24 &34.84 &59.38 &19.72 &22.17 &40.06 &1.18 &1.57 &1.48 \\
 & 6 &19.40 &29.92 &55.91 &17.30 &15.98 &27.34 &1.12 &1.87 &2.04 \\
 & 8 &12.90 &28.93 &47.05 &10.77 &12.64 &19.27 &1.20 &2.29 &2.44 \\

\midrule
\multirow{5}{*}{\rotatebox[origin=c]{80}{\textbf{SPMV}}} &  1 &41.70 &49.48 &73.15 &40.82 &48.06 &72.08 &1.02 &1.03 &1.01 \\
 & 2 &34.47 &40.35 &68.36 &27.14 &37.12 &51.23 &1.27 &1.09 &1.33 \\
 & 4 &32.49 &38.89 &62.46 &25.85 &30.6 &45.11 &1.26 &1.27 &1.38 \\
 & 6 &37.29 &39.24 &55.33 &22.02 &24.66 &38.89 &1.69 &1.59 &1.42 \\
 & 8 &28.42 &34.68 &49.88 &14.78 &19.19 &27.23 &1.92 &1.81 &1.83 \\

\midrule
\multirow{5}{*}{\rotatebox[origin=c]{80}{\textbf{Montage}}} & 1 &68.24 &80.94 &106.43 &66.52 &78.41 &104.87 &1.03 &1.03 &1.01 \\
 & 2 &60.06 &77.13 &97.66 &41.89 &60.81 &76.22 &1.43 &1.27 &1.28 \\
 & 4 &58.00 &76.11 &93.33 &37.89 &47.13 &61.24 &1.53 &1.61 &1.52 \\
 & 6 &61.76 &73.1 &89.38 &22.20 &32.29 &47.89 &2.78 &2.26 &1.87 \\
 & 8 &49.65 &69.29 &85.66 &16.63 &18.45 &33.52 &2.98 &3.76 &2.56 \\   

\midrule
\multirow{5}{*}{\rotatebox[origin=c]{80}{\textbf{\shortstack{Geomean\\ 11 benchm.}}}} &  1 & -- & -- & -- & -- & -- & -- & 1.03 & 1.02 & 1.02 \\          
 &  2 & -- & -- & -- & -- & -- & -- & 1.33 & 1.15 & 1.16 \\          
 &  4 & -- & -- & -- & -- & -- & --  & 1.35 & 1.23 & 1.23 \\          
 &  6 & -- & -- & -- & -- & -- & --  & 1.62 & 1.32 & 1.33 \\          
 &  8 & -- & -- & -- & -- & -- & -- & 1.87 & 1.47 & 1.46 \\  

\bottomrule

\end{tabular}
}
\end{table}

As depicted in Table~\ref{tab:tilesizes}, with too small tile sizes, the communication overhead from more tasks is large enough to slow the benchmark down. However, the performance sharply degrades from~\SI[detect-weight=true]{5}{\kilo{}} to~\SI[detect-weight=true]{7}{\kilo{}} tile sizes for all benchmarks. With too large tile sizes, not enough parallelism takes place in the framework. As most benchmarks see their best performance in the~\SI[detect-weight=true]{5}{\kilo{}} tile size range, all results are presented with a  default tile size of~\SI[detect-weight=true]{5}{\kilo{}}. This comparison uses simulation times, as we have shown that the accuracy of simulation is comparable with execution in Table~\ref{tab:times}.

Figure~\ref{fig:schedule} shows the results of scheduling tasks for the Markov
benchmark with an increasing number of nodes in the network. The memory
allocation tasks \isrc{malloc} are scheduled first on each node followed by
data initialization tasks \isrc{fillzero}. Communication tasks are split 
into \isrc{send} on the sender's side and \isrc{recv} on the receiver's side.
Matrix addition and multiplication take
place in computation tasks \isrc{addmul!}. Finally, \isrc{takecopy!} copies
data from worker nodes into the master for the final matrix computation. 

Figure~\ref{fig:Markov2Node} depicts a network with just the master and one
worker node, and the schedule contains a total of~421 CMM tasks. Communication
is sparse with the low number of nodes, being dominant at the beginning
(sending data to the worker node) and at the end (receiving data from the
worker node). Thus, only two communication processes in the master are 
sufficient to ensure load balance. With more nodes, as depicted in
Figure~\ref{fig:Markov4Node}, two communication processes in the master are 
insufficient to ensure a balanced workload across the network. Processes would
be busy to schedule new communication requests in a timely fashion, and there is a
higher number of tasks to schedule, with 579~CMM tasks in the depicted
schedule. Adding communication processes in the master thereby improves load
balance by allowing communication requests to be finished earlier. However, it
is naive to add an arbitrary number of communicators, as too many causes
network contention by saturating the network with excess requests, worsening
the latency of individual send operations. In such a scenario, the framework
prefers utilizing the node-level tile cache in ``seeded'' nodes over 
sending matrix suboperations to new nodes.
In all schedules, the first tasks on worker nodes are scheduled after the
first tasks on the master node because no communication is needed to schedule on the latter.
Thus, with larger networks, the length of the startup phase is made more prominent.

\begin{table}[b]
\centering
\footnotesize
\renewcommand{\arraystretch}{0.9}
\caption{Speedups for a configuration of eight nodes over one node at a matrix size of~\SI[detect-weight=true]{10}{\kilo{}} and a tile size of~\SI[detect-weight=true]{5}{\kilo{}}.
The one node configuration utilizes the AWS c5.24xlarge instance, which acts as the master in the eight node configuration, with the remaining nodes being AWS c5.9xlarge.
Observed speedups are 
without dynamic tiling. New speedups are calculated with dynamic tiling enabled. Upper bounds assume zero communication time.
All speedups assume that other optimizations, including the automatic communicator configuration, are present.}
\label{tab:speedup}
\begin{tabular}{>{\centering\arraybackslash}p{1.6cm}>{\centering\arraybackslash}p{1.6cm}>{\centering\arraybackslash}p{1.2cm}>{\centering\arraybackslash}p{2.6cm}}
\toprule
{\textbf{Benchmark Name}} &
{\textbf{Observed Speedup (\si{}{$\times$})}} &
{\textbf{Dynamic Tiling (\si{}{$\times$})}} &
{\textbf{Upper bounds (w/o comm.) (\si{}{$\times$})}} \\
\cmidrule{1-4}
    Markov & 3.41 & 4.99 & 5.39 \\
    Kmeans & 3.57 & 4.92 & 5.48 \\
    Hill & 3.32 & 4.13 & 4.71 \\
    Leontief & 2.14 & 2.26 & 3.90 \\
    Synth & 2.48 & 2.86 & 4.23 \\
    Reachability & 1.74 & 2.28 & 3.96 \\
    Hits & 1.84 & 2.33 & 3.95 \\
    BFS & 2.39 & 3.11 & 3.40 \\
    MM & 2.37 & 3.21 & 3.75 \\
    SPMV & 1.55 & 3.35 & 3.84 \\
    Montage & 2.42 & 4.25 & 5.12 \\
\bottomrule
\end{tabular}
\end{table}

\begin{table}
\centering
\renewcommand{\arraystretch}{0.9}
\footnotesize
\caption{Demonstration of the impact of matrix algebraic optimizations on the performance of Julia~CMM, with a matrix size of~\SI[detect-weight=true]{10}{\kilo{}} and a tile size of~\SI[detect-weight=true]{5}{\kilo{}} on a cluster of eight nodes, with the master node using an AWS c5.24x large instance and worker nodes using c5.9xlarge instances. The unoptimized time shows the performance with all algebraic optimizations disabled, the optimized time shows the performance with all optimizations enabled, the speedup column shows the relative speedup gained from algebraic optimizations, the optimizations column shows the total number of algebraic optimizations performed, and the rewriting time column shows the time spent rewriting the expression trees based on the optimizations to be done.
}
\label{tab:algImpact}

{ 
\setlength\tabcolsep{1.0pt}

\begin{tabular}{c|cc|cc|cc|cc|cc}

\toprule

\multirow{3}{*}{\textbf{Name}} &
\multicolumn{2}{|c}{\textbf{Unoptimized}} &
\multicolumn{2}{|c}{\textbf{Optimized}} &
\multicolumn{2}{|c}{\textbf{Speedup (\si{}{$\times$})}} &
\multicolumn{2}{|c}{\textbf{Optimi}} &
\multicolumn{2}{|c}{\textbf{Rewriting}} \\
& 
\multicolumn{2}{|c}{\textbf{Time (\si{\second})}} &
\multicolumn{2}{|c}{\textbf{Time (\si{\second})}} &
\multicolumn{2}{|c}{} &
\multicolumn{2}{|c}{\textbf{zations}} &
\multicolumn{2}{|c}{\textbf{Time (\si{\second})}} \\

\cmidrule{2-11}

 &
\SI{10}{\kilo{}} &
\SI{20}{\kilo{}} &
\SI{10}{\kilo{}} &
\SI{20}{\kilo{}} &
\SI{10}{\kilo{}} &
\SI{20}{\kilo{}} &
\SI{10}{\kilo{}} &
\SI{20}{\kilo{}} &
\SI{10}{\kilo{}} &
\SI{20}{\kilo{}} \\

\midrule
Markov &2.38 &13.10 &2.22 &10.6 &1.07 &1.24 &289 &499 &0.038 &0.078 \\
Kmeans &3.15 &13.72 &2.94 &9.22 &1.07 &1.49 &229 &334 &0.043 &0.077 \\
Hill &3.46 &16.82 &2.73 &11.19 &1.27 &1.50 &197 &309 &0.031 &0.086 \\
Leontief &7.40 &19.52 &6.38 &15.28 &1.16 &1.28 &87 &166 &0.024 &0.055 \\
Synth &4.47 &11.37 &3.6 &9.5 &1.24 &1.20 &110 &230 &0.043 &0.047 \\
Reachability &6.53 &16.07 &5.86 &14.72 &1.11 &1.09 &60 &220 &0.015 &0.084 \\
Hits &6.25 &19.56 &5.42 &16.27 &1.15 &1.20 &108 &245 &0.030 &0.053 \\
BFS &9.29 &22.65 &8.73 &19.67 &1.06 &1.15 &189 &313 &0.058 &0.099 \\
MM &15.93 &25.05 &13.16 &21.72 &1.21 &1.15 &195 &874 &0.064 &0.094 \\
SPMV &17.72 &32.24 &16.19 &27.71 &1.09 &1.16 &105 &672 &0.046 &0.077 \\
Montage &29.06 &42.68 &26.73 &38.86 &1.09 &1.10 &324 &1067 &0.075 &0.122 \\
Average &-- &-- &-- &-- &-- &-- &-- &-- &0.039 &0.076 \\

\bottomrule

\end{tabular}
}
\end{table}

\begin{table}[b]
\centering
\footnotesize
\renewcommand{\arraystretch}{0.9}
\caption{Simulated makespan to run the Markov benchmark with a matrix
size of~\SI[detect-weight=true]{20}{\kilo{}} and a tile size
of~\SI[detect-weight=true]{5}{\kilo{}}, based on the number of communicator and
worker processes in the master node for a 12-node cluster comprised of
a master node (AWS c5.24xlarge instance) and c5.9xlarge instances for
the worker nodes.
The bandwidth
of the c5.24xlarge instance has been limited
to~\SI{10}{\giga\bit\per\second} to remain consistent with the c5.9xlarge 
instances. The number of worker processes in the master node ranges from 2--5,
and the number of communicator processes from 4--14 at a stride of~2.
Durations are averaged over a total of 20~runs.
}
\label{tab:commWorker}
\begin{tabular}{cc|rrrr}
\toprule
& & \multicolumn{4}{c}{\textbf{Simulated Makespan (\si{\second})}} \\
\cmidrule{1-6}
\multicolumn{2}{c|}{\textbf{Num.\ Workers:}} & 2 & 3 & 4 & 5 \\
\midrule
\multirow{6}{*}{\rotatebox[origin=c]{90}{\textbf{Num.\ Comm.}}} & 14 &9.02 &7.85 &6.39 &7.47 \\
& 12 &9.33 &7.97 &6.41 &7.59 \\
& 10 &9.56 &8.26 &6.45 &7.98 \\
& 8 &10.19 &9.13 &6.64 &8.51 \\
& 6 &10.87 &9.98 &6.89 &9.02 \\
& 4 &11.89 &10.34 &7.18 &9.42 \\
\bottomrule
\end{tabular}
\end{table}

\begin{table}[b]
\centering
\footnotesize
\renewcommand{\arraystretch}{0.9}
\caption{The maximum number of communicator processes per maximum bandwidth limit before network
bandwidth saturation is detected, based on the network bandwidth capacities offered
by AWS C5 instances~\cite{AWS:c5}.}
\label{tab:bandwidth}
\begin{tabular}{c|r}
\toprule
{\textbf{Bandwidth (~\SI{}{\giga\bit\per\second} )}} &
{\textbf{Num.\ Comm.}} \\
\midrule
10 &14 \\
15 &18 \\
20 &22 \\
25 &24 \\
\bottomrule
\end{tabular}
\end{table}

Figure~\ref{fig:speedup} depicts the maximum speedups of eleven benchmarks from
one c5.24xlarge AWS instance (96~vCPUs) to fourteen nodes extended with c5.9xlarge AWS instances (564~vCPUs).
Most maximum speedups exceed~\SI{2}{\times},
with some reaching~\SI{5}{\times}. The Markov benchmark reaches a speedup
of~\SI{5.01}{\times} with the best performing tile size, and the Montage
benchmark reaches~\SI{4.34}{\times} maximum speedup compared to vanilla Julia.
For all
configurations, performance improves with dynamic tiling, as it helps to
mitigate the communication bottleneck. Dynamic tiling ensures all workers are
active, and the node-level tile cache ensures they remain active, improving 
performance particularly with higher node counts.

Table~\ref{tab:times} demonstrates the framework's scalability at larger matrix sizes within the same benchmark and network parameters, with comparable results still obtainable at benchmarks of a larger size without significant failure points, as a result of the dynamic tiling optimization encouraging the usage of all nodes in parallelizing larger matrices. A consistent trend amongst all benchmarks is that with increasing network size, the benchmark performance improves as a result. However, as reflected in Figure~\ref{fig:speedup}, the degree of improvement slows down at $\ge$ 10 nodes in most benchmarks when working with a homogeneous cluster, where minimal speedup improvement is obtained from adding more nodes. This is the result of the limited number of communication processes available in the master node, which did not scale up with the network size. Thus, additional nodes compete with existing nodes for receiving data from the master node albeit its limited number of communication processes. The dynamic tiling optimization is ineffective, where after the startup phase in an excessively large network, the master node is unable to efficiently distribute data to all worker nodes without experiencing significant stalls in the worker nodes.

Thus, we present the performance of dynamic tiling and auto-communication configuration on a non-homogeneous cluster where the master node is significantly more powerful than the worker nodes, and thereby contains more communication processes. At higher node counts in the cluster, the performance improvement obtained with the heterogeneous configuration is~$\ge$~\SI{40}{\percent}, and goes up to~\SI{150}{\percent}. Likewise, we find that the declining performance improvement gains with larger network sizes is not as prevalent with a stronger master node, as depicted in Figure~\ref{fig:speedup}. This is the result of the greater communication capacity allowing for more inter-node communication to take place, and when coupled with dynamic tiling, facilitates higher inter-node communication bandwidth. We find that the performance improvement is substantial over the homogeneous cluster, which is to be expected due to the communication being less of a bottleneck.

We note that with heterogeneous networks, both speedups for with and without dynamic tiling generally exhibit a linear pattern with a gradual performance drop at higher node counts. We observed that the speedup improvement gained with dynamic tiling for the tested heterogeneous network is generally not as large as the improvement gained in a homogeneous network of c5.9xlarge AWS instances. However, as the heterogeneous configuration is question was created to mitigate communication limitations, this is to be expected. We note that the dynamic tiling's effectiveness is more apparent at higher node counts which require higher communication costs. Furthermore, the improvement in performance with just the automatic communicator configuration as opposed to vanilla Julia indicates that with minimal user input, the framework is capable of automatically configuring a given cloud network such that it reduces communication latency between potentially disparate nodes. With the inclusion of dynamic tiling, both optimizations help to reduce the effect of communication latency on an application's potential performance.

Table~\ref{tab:dynamic} demonstrates the performance improvement from the dynamic tiling optimization. For $\ge$2 nodes in most benchmarks, it results in a~$\ge$~\SI{10}{\percent} speedup. The improvement is most apparent with Montage at~\SI[detect-weight=true]{276}{\percent}, which is due to it already being structured as a workflow that is flexible to dynamic tile sizes. We observe a general trend that with increasing input matrix size, the performance improvement with the optimization gradually decreases. This can be attributed to the tile size of~\SI{5}{\kilo{}} comprising a smaller portion of a larger matrix, and thus not affecting the overall schedule as much. A large starting tile size in a larger matrix size of~\SI[detect-weight=true]{20}{\kilo{}} results in longer individual tasks to schedule, which does not scale well with the optimization.

Table~\ref{tab:algImpact} depicts the performance gain achieved with all the matrix algebraic optimizations applied during the tree rewriting stage. We find that the speedup achieved is not insignificant, with most speedups being~$\ge$~\SI{10}{\percent} to as high as~\SI{50}{\percent}, thereby demonstrating the effectiveness of the optimizations. We find that the time spent conducting the optimizations at runtime is generally on the order of milliseconds, which is comparatively insignificant to the overall time spent executing the entire benchmark, and thereby does not adversely affect the performance.

Table~\ref{tab:commWorker} demonstrates the time it takes to run a Markov benchmark with varying configurations of communicator and worker processes in the master node. We find that there is a trend of worsening performance with reducing the number of communicators while keeping the same number of workers, which is expected as less data is able to be propagated to worker nodes in the same amount of time. We find that increasing the number of workers usually improves the performance then degrades after a certain point. This occurs by introducing too many worker processes in the master node, causing the framework to favor scheduling tasks there due to its zero communication cost. Finally, we find that allocating less communicators with more workers can achieve a higher performance result than allocating the maximum number of communicator processes permissible by the bandwidth limit, and then filling the remaining cores with worker processes.
We find that when working with a~\SI{10}{\giga\bit\per\second} bandwidth network, we can allocate up to 14 communicators in the master node before network saturation is detected, as indicated in Table~\ref{tab:bandwidth}. In Table~\ref{tab:commWorker}, the performance achieved with 2 or 3 worker processes in the master is consistently weaker than with 4 worker processes, even when allocating the maximum number of communicators. This demonstrates that focusing only on communication awareness is naive, and the availability of computational resources should additionally be taken into account. Thus, instead of adhering to simply to maximizing communication capability, we aim to balance the two amounts according to the resources available on the master node.

By Amdahl's Law, we calculate the upper bound of the speedups that is attainable with an infinitely fast network by setting communication latency to zero in the simulation. 
Other parameters remain the same: eight nodes with a matrix size of~\SI[detect-weight=true]{10}{\kilo{}} and tile size of~\SI[detect-weight=true]{5}{\kilo{}}. 
As all tasks are parallelizable and asynchronous, communication acts as the
bottleneck to true parallelism.  Table~\ref{tab:speedup} shows that with
dynamic tiling, most benchmarks do not significantly deviate from the
upper bound, being in the range
of~\SIrange[range-units=single,range-phrase=--]{9}{43}{\percent} from the
upper bound.  This is expected, as the upper bound is practically infeasible
due to necessitating instantaneous communication.

\section{Conclusions}
We proposed the CMM framework which extends the Julia language to automatically parallelize matrix computations for the cloud with minimal user intervention. The framework
employs offline profiling to generate accurate time prediction models using
polynomial regression. They are used with tiled dependencies by the modified HEFT scheduler to allocate tasks such that
overall execution time is minimized. Communication overhead optimizations are made by introducing additional
communication processes, dynamic tiling, and the node-level tile cache. The implementation of an automatic communication configuration allowing the transition to heterogeneous clusters led to improvements in reducing the effect of communication overhead. We
conducted an extensive experimental evaluation on a set of benchmarks
using up to fourteen nodes (564~\makebox{vCPUs}) in the AWS public cloud. Our
framework achieved speedups of up to a factor of~\SI{5.1}{}, with an average~\SI{74.39}{\percent} of the upper bound for speedups.

\clearpage
\bibliographystyle{ACM-Reference-Format}
\bibliography{references}

\end{document}